\documentclass[preprint, 12pt]{elsarticle}

\usepackage{amssymb}
\usepackage{amsthm}
\usepackage{subcaption}
\usepackage{algorithm}
\usepackage{algpseudocode}
\usepackage{hyperref} 
\usepackage{adjustbox}
\usepackage[shortcuts,acronym,nonumberlist,nogroupskip]{glossaries}
\usepackage{nomencl}
\usepackage{etoolbox}
\usepackage{listings} 
\usepackage{jlcode} 
\usepackage{pifont}
\newcommand{\cmark}{\ding{51}}%
\newcommand{\xmark}{\ding{55}}%

\makeglossaries 
\newacronym{am}{AM}{active material loss}
\newacronym{bm}{BM}{bucket model}
\newacronym{bms}{BMS}{battery management system}
\newacronym{bess}{BESS}{battery energy storage system}
\newacronym{der}{DER}{distributed energy resources}
\newacronym{dla}{DLA}{Direct Lookahead}
\newacronym{ems}{EMS}{energy management systems}
\newacronym{ema}{EMA}{energy management algorithm}
\newacronym{ess}{ESS}{energy storage systems}
\newacronym{ecm}{ECM}{equivalent circuit model}
\newacronym{ev}{EV}{electric vehicle}
\newacronym{fom}{FOM}{full order model}
\newacronym{hess}{HESS}{hybrid energy storage system}
\newacronym{hp}{HP}{heat pump}
\newacronym{nmc}{NMC}{nickel manganese cobalt oxides}
\newacronym{mpc}{MPC}{Model Predictive Control}
\newacronym{empc}{eMPC}{economic Model Predictive Control}
\newacronym{mces}{MCES}{multicarrier energy systems}
\newacronym{ocv}{OCV}{open-circuit voltage} 
\newacronym{ocp}{OCP}{optimal control problem} 
\newacronym{pei}{PEI}{power electronic interface}
\newacronym{pb}{PB}{physics-based}
\newacronym{pbrom}{PBROM}{physics-based reduced order model}
\newacronym{rl}{RL}{reinforcement learning}
\newacronym{scm}{SCM}{series connected module}
\newacronym{sei}{SEI}{solid electrolyte interface}
\newacronym{sdp}{SPD}{sequential decision problem}
\newacronym{soc}{SOC}{State of Charge}
\newacronym{soh}{SOH}{State of Health}
\newacronym{spv}{SPV}{solar photovoltaics}
\newacronym{st}{ST}{solar thermal}
\newacronym{tess}{TESS}{thermal energy storage system}
\newacronym{lib}{LIB}{Li-ion batteries}
\newacronym{lto}{LTO}{Lithium-titanate}
\newacronym{lfp}{LFP}{Lithium iron phosphate}
\newacronym{umf}{UMF}{Universal Modeling Framework}

\makenomenclature
\renewcommand\nomgroup[1]{%
  \item[\bfseries
    \ifstrequal{#1}{V}{Physics constants}{%
    \ifstrequal{#1}{N}{Number sets}{%
    \ifstrequal{#1}{A}{Optimization, Control \& Dynamics}{%
    \ifstrequal{#1}{C}{Battery Performance Model}{%
    \ifstrequal{#1}{D}{Battery Degradation Model}{%
    \ifstrequal{#1}{E}{Electric Vehicle (EV) Model}{
    }}}}}}%
]}

\nomenclature[A]{$w=[w_{\text{grid}}, w_{\text{loss}}, w_{\text{SoC}}, w_{\text{TESS}}]$}{Objective function weights.}
\nomenclature[A]{$\tilde{\ \ }$}{Approximated/estimated variable/function.}
\nomenclature[A]{$\overline{\ \ }, \underline{\ \ }$}{Maximum and minimum variable bounds.}
\nomenclature[A]{$^{+}, ^{-}$}{Positive and negative variable components.}
\nomenclature[A]{$\pi$}{policy}
\nomenclature[A]{$t, t', \Delta t$}{Time index, time index within a policy $\pi$ and timestep.}
\nomenclature[A]{$\mathbb{E}_{W}$}{Expectation with respect to the exogenous process $W$}
\nomenclature[A]{$a \in \mathbb{A}$}{Set of energy assets.}
\nomenclature[A]{$b \in\{\text{BESS}, \text{EV}\} \subset a$ }{Battery storage asset index}
\nomenclature[A]{$X^{\pi}_t(\cdot)$}{Policy function at time $t$}
\nomenclature[A]{$P_{a,t} \in \mathcal{P}$}{Power setpoint for system $a$ at time $t$}
\nomenclature[A]{$\mathcal{P}_{a,[t,t+H]}$}{Power trajectory for system $a$ over horizon $[t,t+H]$}
\nomenclature[A]{$\mathcal{S}_{a,[t,t+H]}$}{State trajectory for system $a$ over horizon $[t,t+H]$}
\nomenclature[A]{$C_{\textrm{grid}}$}{Cost of electricity from the grid}
\nomenclature[A]{$C_{\textrm{loss}}$}{Loss cost or penalty}
\nomenclature[A]{$S_{a,t} \in \mathcal{S}$}{State of system $a$ at time $t$}
\nomenclature[A]{$S^M_{a,t}(\cdot)$}{Transition function for asset $a$ at time $t$}
\nomenclature[A]{$\theta_{a,t}$}{Model parameters for system $a$ at time $t$}
\nomenclature[A]{$W_{t+1}$}{Exogenous information process at time $t+1$}
\nomenclature[A]{$B_{a,t}$}{Belief state for asset $a$ at time $t$}
\nomenclature[A]{$p_{\textrm{SoCDep}}$}{Penalty term for SoC deviation}
\nomenclature[A]{$p_{\textrm{TESS}}$}{Soft-constraint penalty of TESS overcharge}
\nomenclature[A]{$\varepsilon_{SoC}$}{Penalty deviation from required departure $SoC$}
\nomenclature[A]{$\lambda_{\text{buy/sell},t}$}{Day-ahead buy and sell prices.}
\nomenclature[A]{$T, H$}{Simulation and optimization horizons.}
\nomenclature[A]{$\mathcal{D}_t^{\pi}=[t;t+H]$}{Prediction horizon window at time $t$}

\nomenclature[E]{$\gamma_{\text{EV},t}$}{EV availability at time $t$ (1 = parked, 0 = driving)}
\nomenclature[E]{$t_{\text{dep/arr}} \sim $$\mathcal{T}_{\text{dep/arr}}$}{Departure and arrival time of the EV}
\nomenclature[E]{$P_{\textrm{tot},\textrm{EV},t}$}{Total power associated with EV at time $t$}
\nomenclature[E]{$P_{\textrm{EV},t}$}{EV charger power at time $t$}
\nomenclature[E]{$P_{\text{drive},\text{EV}}$}{Power consumed while driving}
\nomenclature[E]{$\mu_{\text{drive}}$}{Mean of EV driving power distribution}
\nomenclature[E]{$\sigma_{\text{drive}}^2$}{Variance of EV driving power distribution}
\nomenclature[E]{$SoC_{\textrm{dep}}^*$}{Required SoC at departure}

\nomenclature[C]{$x_{b,t}=i_{b,t}$}{Control cell current to the battery model at time $t$}
\nomenclature[C]{$SoC_{b,t}$}{State of Charge of battery $b$ at time $t$}
\nomenclature[C]{$\tilde{S}^M_{b,t}(\cdot)$}{Battery transition model with performance and ageing}
\nomenclature[C]{$p^M_{b,t}(\cdot)$}{Battery performance model}
\nomenclature[C]{$v_{t,b,t}$}{Terminal voltage of battery $b$}
\nomenclature[C]{$S_{b,t}=[SoC, v_{t}]^T_{b,t}$}{Battery state vector at time $t$ }
\nomenclature[C]{$d^M_{b,t}(\cdot)$}{Battery ageing model}
\nomenclature[C]{$t_{0,b}$}{Battery elapsed lifetime}
\nomenclature[C]{$N_{s,b}$}{Number of cells in series}
\nomenclature[C]{$N_{p,b}$}{Number of parallel branches}
\nomenclature[C]{$\eta_c$}{Coulombic efficiency}
\nomenclature[C]{$Q_{b,t}$}{Battery capacity at time $t$}
\nomenclature[D]{$SoH$}{State of Health of the battery}
\nomenclature[D]{$i_{\text{cycle},b,t}$}{Cyclic degradation current}
\nomenclature[D]{$i_{\text{cal},b,t}$}{Calendar degradation current}
\nomenclature[D]{$i_{\text{loss},b,t}$}{Total degradation current}
\nomenclature[D]{$i_{\textrm{SEI},b,t}$}{SEI side reaction current}
\nomenclature[D]{$\eta_{k,b,t}$}{SEI kinetic overpotential}
\nomenclature[C]{$OCV_{b,t}$}{Open-circuit voltage at time $t$}
\nomenclature[D]{$OCV_{n,b,t}$}{Open-circuit voltage of anode}
\nomenclature[D]{$z_{b,t}, z_{b,0\%}, z_{b,100\%}$}{Lithium stoichiometry at time $t$, and limits at 0 and 100\%}
\nomenclature[D]{$i_{\textrm{AM},b,t}$}{Active material loss current}
\nomenclature[D]{$\delta_{\textrm{SEI},b,t}$}{SEI layer thickness}
\nomenclature[D]{$\varepsilon_{e,b,t}$}{Electrolyte volume fraction}

\journal{Journal of Energy Storage}

\begin{document}

\begin{frontmatter}



\title{Ageing-aware Energy Management for Residential Multi-Carrier Energy Systems}


\author[inst1]{Darío Slaifstein}
\author[inst1]{Gautham Ram Chandra Mouli}
\author[inst1]{Laura Ramirez-Elizondo}
\author[inst1]{Pavol Bauer}

\affiliation[inst1]{organization={DC Systems, Energy Conversion \& Storage, Electrical Sustainable Energy Department, Delft University of Technology},
            addressline={Mekelweg 8}, 
            city={Delft},
            postcode={2628}, 
            state={Zuid-Holland},
            country={Netherlands}}

\begin{abstract}
In the context of building electrification, the operation of distributed energy resources integrating multiple energy carriers (electricity, heat, mobility) poses a significant challenge due to the nonlinear device dynamics, uncertainty, and computational issues. As such, energy management systems seek to decide the power dispatch in the best way possible. The objective is to minimize and balance operative costs (energy bills or asset degradation) with user requirements (mobility, heating, etc.). Current energy management uses empirical battery ageing models outside of their specific fitting conditions, resulting in inaccuracies and poor performance. Moreover, the link to thermal systems is also overlooked. This paper presents an ageing-aware nonlinear economic model predictive controller for electrified buildings that incorporates physics-based battery ageing models. The models distinguish between energy storage systems (chemistry, ageing state, etc.) and make explicit the trade-off between grid cost and battery degradation. The proposed algorithm can either cut down on grid costs or extend battery lifetime (electric vehicle or stationary battery packs). Additionally, substituting NMC cells with LFP chemistries optimizes grid performance during the summer, yielding a 10\% grid cost reduction and a 20\% decrease in degradation. Finally, the grid cost and degradation of the presented MPC when using aged batteries are improved with respect to the state of the art by 10\% and 5\% respectively, in periods with high solar generation and low thermal loads like summer.
\end{abstract}

\begin{graphicalabstract}
\centering
\includegraphics[width=1.2\linewidth]{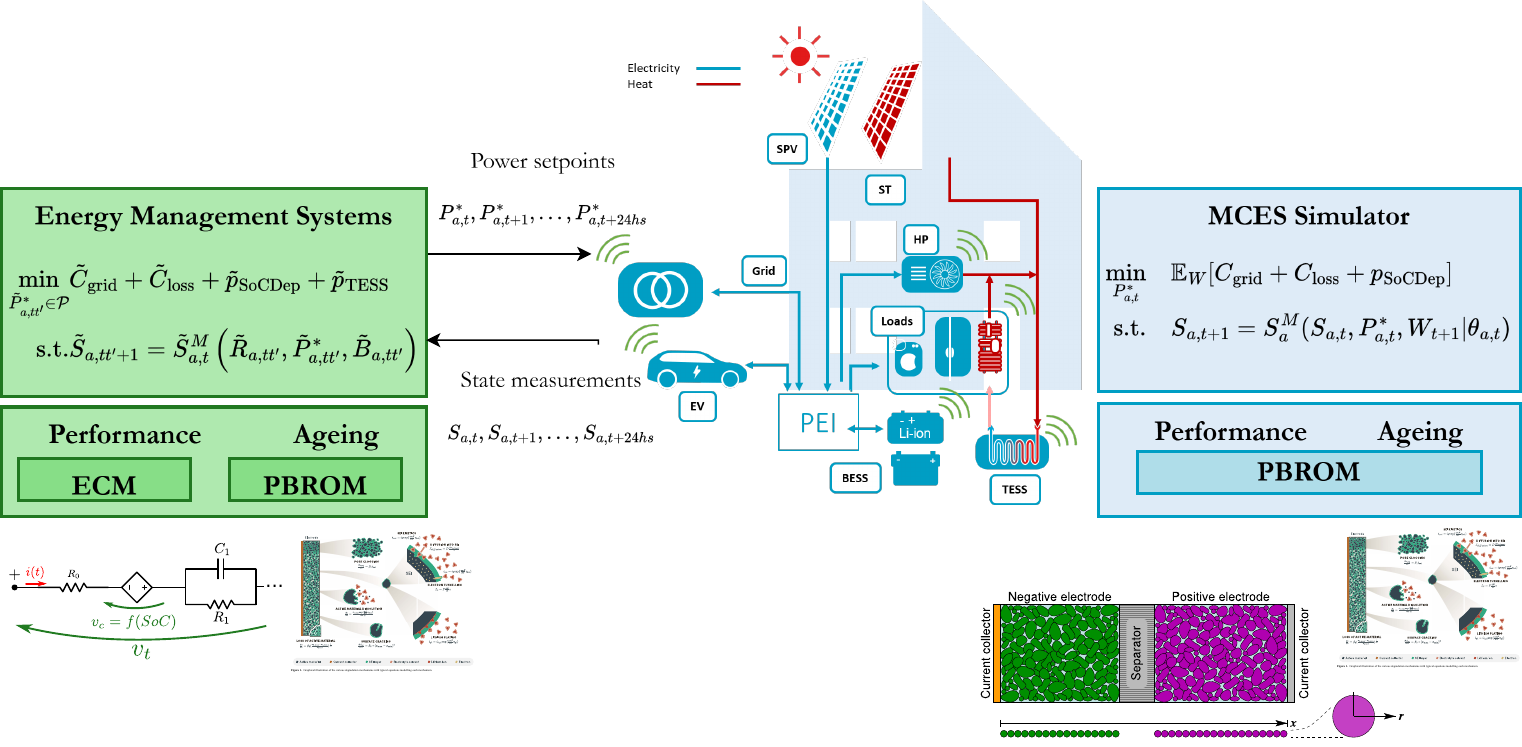}
\end{graphicalabstract}

\begin{highlights}
\item Integrated physics-based degradation models into optimal predictive energy management for multi-carrier buildings. This allows users to trade-off between grid cost reductions and battery lifetime extension.
\item Novel algorithm can distinguish between cathode chemistries, with LFP cells achieving lower grid costs and capacity fade than their NMC counterpart.
\item Integrating physics-based ageing improves EMS performance with aged batteries, while traditional benchmarks increase their grid cost and capacity fade.
\end{highlights}


\end{frontmatter}


\section{Introduction}
\label{sec:intro}
The decarbonization of the economy as a whole is a significant challenge for modern societies. In particular, the sustainable transformation of both the energy and transport sectors poses significant technical and cultural challenges \cite{IEA2021}. Both transitions couple in the population’s homes where electricity, mobility, or heat are needed. Thus, possible synergies between the three systems can be exploited to achieve the desired decarbonization, freedom, resiliency, and cost savings at the local or aggregated level \cite{Geidl2007}. The successful exploitation of such coupling thus needs to be carefully tailored and built into the design of modern multi-carrier energy systems \cite{Andersson2005, Geidl2007, Vermeer2020, Vermeer2022, Ceusters2021, Ceusters2023, VanDerMeer2018, Ye2020}. This necessarily leads to advanced \ac{ems} that schedule and control the \ac{der} \cite{Vermeer2022b, Vermeer2022AData, Ceusters2021,Ye2020, EsmaeelNezhad2022}. Thus, the \ac{ems} needs to handle uncertainty introduced by electric vehicles \cite{Alexeenko2023AchievingStudy}, solar generation and loads, as well as the battery degradation \cite{Xavier2021, Jin2022, Li2019, Plett2015, Plett2016}. The main goal of this work is to address this last point.

\begin{figure}[tb]
    \centering
    \includegraphics[width=1.15\textwidth]{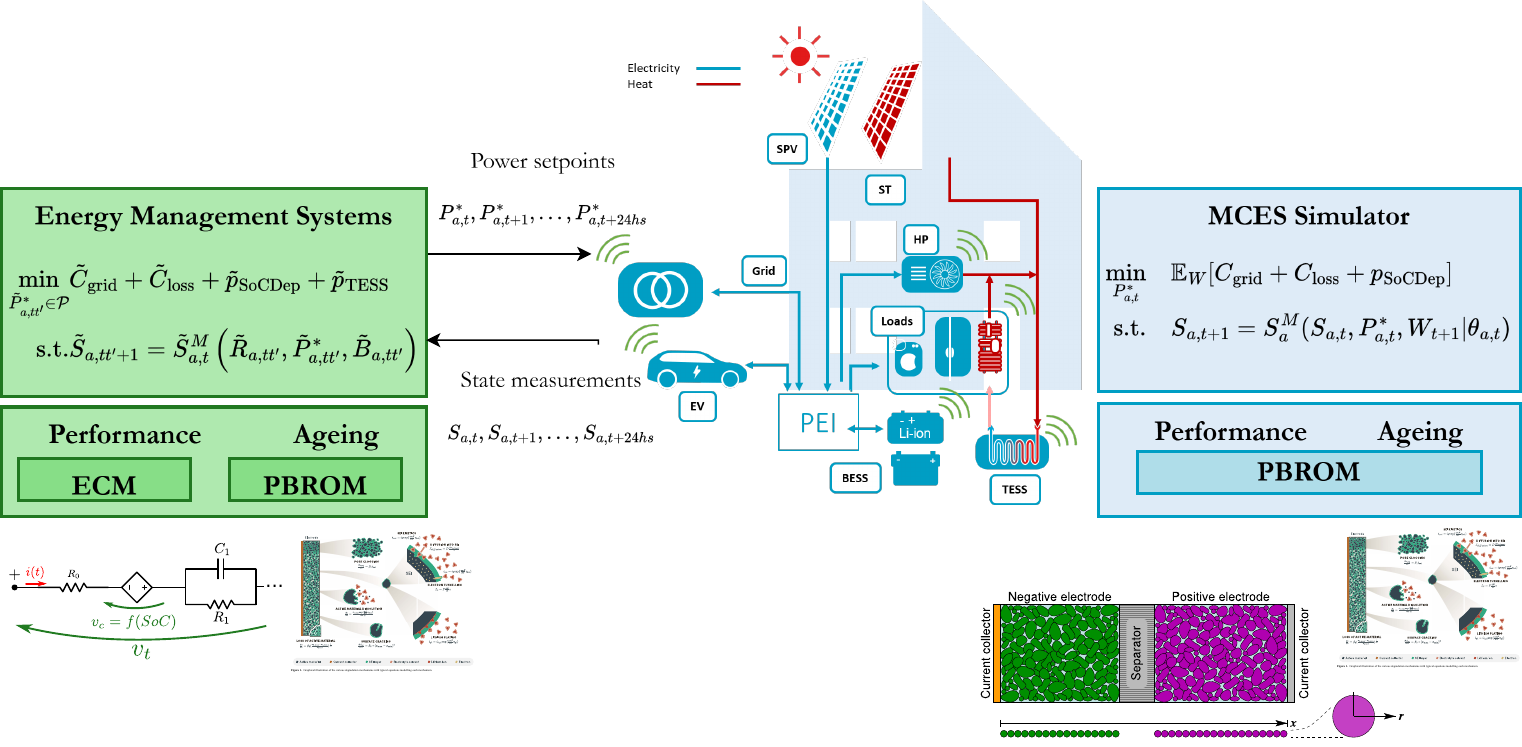}
	\caption{Schematic diagram of the proposed electrified multi-carrier building.}
	\label{fig:FLXconcept}
\end{figure}

To dispatch and operate residential \ac{mces}, the literature suggests \ac{mpc} \cite{EsmaeelNezhad2022, Ceusters2021}, stochastic optimization \cite{Chen2012}, \ac{rl} \cite{Ye2020,Ceusters2021,Ceusters2023} and many others. Usually, the basis of such advanced systems is a day-ahead plan or dispatch that schedules the power of the assets along the day \cite{Risbeck2018Mixed-IntegerSystems, Vermeer2020, Vermeer2022b, Vermeer2022AData, Jouini2024PredictiveStudy, Li2021IntradaySystem}. This planner is usually an optimization-based system that uses approximated deterministic forecasts of certain inputs to schedule the different assets. The decisions taken are then implemented and modified in real time. The optimization models have to model the representative aspects of the different assets of the energy system. This includes their power limits, dynamics, and operational constraints. The standard approach is to limit the models to simplified linear or quadratic forms, overlooking most technological particularities \cite{Geidl2007, Ye2020, Ceusters2021, Mariano-Hernandez2021ADiagnosis, Yang2023BuildingPricing, Mittelviefhaus2022ElectrificationMitigation, Su2025DynamicSystems, Karthikeyan2025EnhancingSystems}. This nonlinearities can be the efficiency of a heat-pump, the voltage response of a battery or the heat losses of a building. In particular, when several energy storage systems are present, the \ac{ems} has to decide which storage system to use and when. In this regard, battery \ac{ess} are limited by their ageing \cite{Jin2022,Li2019,Plett2015,Plett2016,Reniers2019,Vermeer2022,Vermeer2022b} or their availability for mobility \cite{Vermeer2022,Alexeenko2023AchievingStudy}. Current \ac{bess} control decisions impact its remaining lifetime, thus, having a control-oriented predictive model that accurately reflects the cost of current decisions is key for optimal operation \cite{VegaGarita2025TheReview}. The bigger the bias between the optimization model and reality the less trustworthy are the decisions and predictions made. 

On the other hand, the interaction of electricity and heat is becoming more relevant as heating electrification intensifies, and single-carrier optimization might lead to under performance and cost inefficiencies. When investigating \ac{mces}, Ceusters et al. \cite{Ceusters2021, Ceusters2023} use first-order linear models for both \ac{bess} and \ac{tess}, neglecting any differences between their dynamics.  Similarly, Ye et al. \cite{Ye2020} does not mention any difference between storage systems nor include \ac{ev}s in their system. Alpizar-Castillo, et al \cite{Alpizar-Castillo2024ModellingHouse} focuses on thermal dynamics, only incorporating \ac{bess} with linear models, without including \ac{ev}s. Other works only focus on the electrical carrier without including coupling electrical heating and storage \cite{Vermeer2020, Vermeer2022b, Li2019, Li2023EnsembleManagement}. These last works apply different battery models to describe their key variables, such as state-of-charge $SoC$, terminal voltage $v_t$, and state-of-health $SoH$. In this multi-carrier context, the optimization needs accurate cost functions and predictive models to decide in which carrier to store energy. If no dynamic distinction is made between \ac{ess} then the predicted operation will be far from reality leading to over investment on inefficient technologies. Thus, this work integrates electrical and thermal storage models together in the same optimization, but using different dynamic models to reflect their particular technologies accordingly. 

Unfortunately, although physical ageing mechanisms have been studied and modeled \cite{Okane2022, Jin2017, Xavier2021, Prada2013Simulations, Wang2014, Vermeer2022}, they have been partially incorporated into \ac{ems} design of residential \ac{mces} through empirical ageing models  \cite{Vermeer2022,Vermeer2022b, Mariano-Hernandez2021ADiagnosis,Risbeck2018Mixed-IntegerSystems}. Battery-ageing models fall within two categories: empirical or \ac{pb} \cite{Plett2015, Vermeer2022, Reniers2019}.  The first are the most widely used in the literature due to their simplicity.  They are obtained by performing long standardized calendar and cycling ageing tests \cite{Reniers2019, Vermeer2022}. Unfortunately, empirical degradation models only have interpolation capabilities, usually use non-linear equations, can only predict regular cycles (average C-rate, minimum $SoC$, etc.), are prone to overfitting, and are chemistry dependent \cite{Wang2014, Schmalstieg2014, Reniers2019}. On the other hand, \ac{pb} models are built through first-principles and specialized tests to identify individual degradation mechanisms \cite{Okane2022, Prada2013Simulations, Reniers2019}. They have extrapolation features, can be expressed in the state-space form,  account for several cathode chemistries, and represent dynamic profiles such as \ac{ev}s, utility scale \ac{bess} or any irregular load. However, they are also non-linear and, in general, non-convex \cite{Okane2022, Jin2017, Purewal2014DegradationModel, Reniers2019, Prada2013Simulations}. The integration of \ac{pb} ageing models into the operation of \ac{bess} has been recently studied at the \ac{bms} level for standalone and \ac{ev} applications \cite{Xavier2021, Reniers2023, Reniers2021, Jin2022, Li2019, Li2024NonlinearCharging} usually through control-oriented \ac{pbrom}. In the cited references, important cost savings were achieved either by preserving battery lifetime or by making an explicit trade-off between the grid costs and capacity fade, even when implementing optimization horizons of a day or less. To the author's best knowledge, their integration into applications where the \ac{bess} interacts with more assets, such as transmission grids, microgrids, industry, and in particular buildings, has not been extensively researched yet \cite{VegaGarita2025TheReview}. In this work,  physics-based models unlock a new level of \ac{ems} intelligence: the ability to differentiate \textbf{unique battery chemistries} and \textbf{evolve alongside the cell's actual aging process}, ensuring optimal dispatch across the entire lifecycle of the multi-carrier system.

\begin{table*}[h]
    \centering
    \caption{Summary of Literature Review.}
    \label{tab:litRev}
    \begin{adjustbox}{width=1.1\textwidth}
    \begin{tabular}{l l | c c c |c c c c| c c ccc}
        \hline
        & & \multicolumn{3}{c}{\textbf{Loads / Carriers}} & \multicolumn{4}{c}{\textbf{Flexibility}} & \multicolumn{5}{c}{\textbf{Battery modelling}}\\
        \textbf{Ref.}& \textbf{Application} & \begin{tabular}[c]{@{}l@{}}\textbf{Elec.}\\ \textbf{Load}\end{tabular} & \begin{tabular}[c]{@{}l@{}}\textbf{Space}\\ \textbf{Heating}\end{tabular} & \textbf{Nat.}\textbf{Gas}& \textbf{TESS} & \textbf{EV} & \textbf{BESS} & \begin{tabular}[c]{@{}l@{}}\textbf{Heat}\\ \textbf{Pump}\end{tabular}& \textbf{Performance}& \textbf{Degradation}&  \begin{tabular}[c]{@{}l@{}}\textbf{Dynamic}\\ \textbf{Profile}\end{tabular} & \begin{tabular}[c]{@{}l@{}}\textbf{Cathode}\\ \textbf{Chemistry}\end{tabular} & \begin{tabular}[c]{@{}l@{}}\textbf{Used and}\\ \textbf{New Cells}\end{tabular}\\
        \hline
        \cite{Ceusters2021, Ceusters2023, Ceusters2023AnSystems} & Energy Hub& \cmark & \cmark & \cmark & \cmark & & \cmark & \cmark & Coulomb counting& &  &&\\
        \cite{Vermeer2022b}& HEMS& \cmark& Flow-based& & & \cmark& \cmark& & Coulomb counting& Empirical&  &&\\
        \cite{Geidl2007}& Energy Hub& \cmark& \cmark& \cmark& \cmark& & \cmark& & Coulomb counting& & & &\\
        \cite{Ye2020} & HEMS& \cmark& \cmark& \cmark& \cmark& & \cmark& & Coulomb counting& & & &\\
        \cite{Yang2023BuildingPricing} & HEMS& \cmark& \cmark& \cmark& \cmark& \cmark& \cmark& & Coulomb counting& & & &\\
        \cite{Karthikeyan2025EnhancingSystems} & Energy Hub& \cmark& & & & \cmark& \cmark& & Coulomb counting& Empirical& & &\\
        \cite{Su2025DynamicSystems} & Energy Hub& \cmark& Flow-based& \cmark& \cmark& & \cmark& \cmark& Coulomb counting& Empirical& & &\\
        \cite{Reniers2021, Reniers2018} & Utility scale& & & & & & \cmark& & \multicolumn{2}{c}{PBROM}& \cmark& *&*\\
        \cite{Xavier2021} & Fast-charging& & & & & \cmark& & & \multicolumn{2}{c}{PBROM}& \cmark& *&*\\
        \cite{Jin2022} & Fast-charging& & & & & \cmark& & & ECM& PBROM& \cmark& *&\cmark\\
        \cite{Li2019} & HEMS& \cmark& & & & & \cmark& & \multicolumn{2}{c}{PB ECM}& \cmark& *&*\\
        \cite{Li2023EnsembleManagement}& HEMS& \cmark& & & & & \cmark& & \multicolumn{2}{c}{PB ECM}& \cmark& *&*\\
        \cite{Dorronsoro2025BatteryModels}& HEMS& \cmark& & & & & \cmark& & ECM&Empirical& & &\cmark\\
        \cite{Park2023OptimalApproach}& Microgrid & \cmark& & & & & \cmark& & Coulomb Counting& Empirical& & &\\
        \textbf{This work} & HEMS & \cmark & \cmark & & \cmark & \cmark & \cmark & \cmark & ECM& PBROM&  \cmark&\cmark&\cmark\\
        \hline
        \multicolumn{14}{l}{Note *: Capable, but no Case Study}\\
    \end{tabular}
    \end{adjustbox}
\end{table*}

The most relevant works can be found in Table \ref{tab:litRev}. In summary, current optimization-based approaches that do use \ac{pbrom} to actively trade off between degradation and economic benefits are restricted to standalone battery systems \cite{Xavier2021, Jin2017, Jin2022, Reniers2018, Reniers2021, Reniers2023, Li2024NonlinearCharging}. When the battery is integrated with larger systems (\ac{spv}, \ac{ev}, microgrid, etc.), battery dynamics are usually simplified and empirical models are used in most literature, leading to inaccurate degradation predictions and under utilization \cite{Vermeer2022b, Li2019, Li2023EnsembleManagement, Dorronsoro2025BatteryModels, Park2023OptimalApproach}. None of the above references include the integration with a thermal carrier or multiple storage devices, and if they do, they do not include ageing models, \cite{Su2025DynamicSystems, Yang2023BuildingPricing, Geidl2007, Ceusters2023, Karthikeyan2025EnhancingSystems}. 



This paper bridges the previously mentioned gaps by introducing an innovative economic MPC framework that integrates \textbf{ageing physics-based reduced-order models (PBROM)} directly into the \ac{ems} algorithm of multi-carrier systems. Current Energy Management Systems (EMS) for residential multi-carrier energy systems (MCES) often operate without a complete understanding of how their decisions impact the long-term health and lifetime of their storage assets. This integration fundamentally changes how the \ac{ems} operates, moving from a short-sighted, price-based optimization to a more holistic, ageing-aware strategy. Our approach enables the system to choose the optimal power dispatch of each storage system based on:
\begin{itemize}
    \item The co-optimization of both \textbf{fast electrical storage} (BESS and EV) and \textbf{slower thermal storage (TESS)}, recognizing their distinct response times and efficiencies, through distinct terminal sets. 
    \item The recognition of different \textbf{cathode chemistries} in lithium-ion batteries, leading to different power dispatches and grid costs.
    \item     The exploitation of parameter differences between \textbf{new and used batteries}, ensuring that the operational plan remains accurate and effective throughout the entire lifespan of the battery storage units.
    \item The capacity to identify \textbf{dominant degradation mechanisms}, which provides the \ac{ems} with the critical information needed to make decisions that not only meet demand but also actively preserve the health of the batteries depending on their ageing state. 
\end{itemize}

\section{Modeling and Optimal Planner}
\label{sec:modeling}

Our \ac{ems} is an optimization-based secondary controller that minimizes energy cost and battery ageing. A schematic of  the \ac{mces} and the \ac{ems} is presented in Fig.~\ref{fig:FLXconcept}. The system is composed of \ac{spv}, \ac{bess}, \ac{ev}, \ac{pei}, \ac{hp}, \ac{st}, \ac{tess}, grid connection and loads. On the left, the \ac{ems} decides the power dispatch $P_{a,t}^*$ at $t \in \mathcal{D}_t$ where $\mathcal{D}_t=[0,T]$, passing it down to the \ac{mces} simulator. The \ac{mces} simulator feeds back the state measurements $S_{a,t}$ to continue with the loop. 

 The following section describes the \ac{ems} models, following the \ac{umf} by Powell \cite{Powell2022}. For a given system size, the objective is to handle the operation cost, which is composed of three parts: the net cost of energy from the grid $C_{\textrm{grid}}$, the degradation cost of losing storage capacity $C_{\textrm{loss}}$, and a penalty for not charging the \ac{ev} $p_{\textrm{SoCDep}}$. The grid cost and the degradation cost are cumulative objectives because the goal is to optimize them through time, while the penalty for not charging the \ac{ev} to the desired $SoC$ level is only a point reward at departure times $t_{\textrm{dep}}$. The \ac{sdp} is then:

\begin{subequations}\label{eq:sdp}
    \begin{align}
        \min_{P^*_{a,t}} \quad & \mathbb{E}_{W}[ C_{\textrm{grid}}+C_{\textrm{loss}}+p_{\textrm{SoCDep}}]\\
        \textrm{s.t.} \quad & S_{a,t+1}=S^M_a (S_{a,t}, P^*_{a,t}, W_{t+1}|\theta_{a,t}) \\
                    & P_{a,t}^*=X^{\pi}_t(S_{a,t}) \in \mathcal{P} & \forall a \in \mathbb{A} & ,\ t \in \mathcal{D}_t \\
                    & S_{a,t} \in \mathcal{S} & \forall a \in \mathbb{A} & ,\ t \in \mathcal{D}_t
    \end{align}
\end{subequations}

\noindent{with}
\begin{equation}
    \mathbb{A}=\{\text{SPV}, \text{grid}, \text{EV}, \text{BESS}, \text{HP}, \text{ST}, \text{TESS}\}\,.
\end{equation}
where the components of the objective are:
\begin{subequations}\label{eq:cgrid}
    \begin{align}
        & C_{\textrm{grid}}=w_{\text{grid}}\sum_{t=0}^{T}{\left(\lambda_{\textrm{buy},t} .P^{+}_{\textrm{grid},t}-\lambda_{\textrm{sell},t} .P^{-}_{\textrm{grid},t}\right).\Delta t} \\
    & C_{\text{loss}}=w_{\text{loss}}.c_{\text{loss}}.\sum_{t=0}^{T}\sum_{b}{N_{s,b} . N_{p,b} . i_{\text{loss},b,t}.\Delta t},\ \forall\ b \in\{\text{BESS}, \text{EV}\}  \subset a, \\
    & p_{\textrm{SoCDep}}=w_{\textrm{SoC}}.||\varepsilon_{\textrm{SoC}, t_{\text{dep}}}||_2^2
    \end{align}
\end{subequations}

\noindent{where $S_{a,t}$ is the state vector, $P^*_{a,t}$ is the optimal decision for timestep $t$, $W_{t+1}$ and is an exogenous process that introduces new information after making a decision. The mappings $S_{a,t}^M(.)$, and $X^{\pi}_t(.)$ are the transition function and optimal policy, respectively. The first is a set of equations describing the states and parameter evolution, and the second is the algorithm that finds the setpoints. The vector $\theta_{a,t}$ contains all the parameters of each asset $a$ and changes over time $t$. The subindex $a \in \mathbb{A}$ corresponds to the assets shown in Fig. \ref{fig:FLXconcept}. The index $b$ denotes the electric storage assets.  The evaluation/simulation time window is $\mathcal{D}_t \in [0,T]$ and the timestep $\Delta t=15$min. $C_{\textrm{loss}}$ is explained in Section \ref{sse:degModel} and the penalty $p_{\textrm{SoCDep}}$ in Section \ref{sse:ev}.}

The following definitions of the elements are considered:
\begin{itemize}
    \item{The actions or decision variables are
        \begin{equation}
        P^*_{a,t}=[P_{\text{EV}}, P_{\text{BESS}}, P_{\text{HP}}^{\textrm{e}}]_t^T\,.
        \end{equation}
    }
    \item{The exogenous processes/inputs to the optimization $W_{t+1}$ are the prices $\lambda$, \ac{ev} availability $\gamma$, the solar power $P_{\text{PV/ST}}$, the electric and the thermal demands $P_{\text{load}}^{\text{e/th}}$:
        \begin{equation}
            W_{t+1}=[\lambda_{\text{buy/sell}}, \gamma_{n_{\textrm{EV}}}, P_{\text{PV}}, P_{\text{ST}}, P_{\text{load}}^{\text{e}}, P_{\text{load}}^{\text{th}}]_{t+1}^T
        \end{equation}
    }
    \item{The state vector has 2 components, the physical state of the system $R_t$, and beliefs about uncertain quantities or parameters $B_t$. All the observable physical quantities of our system, such as currents, voltages, and so on, are included in $R_t$. Finally, our belief state $B_{a,t}$ is composed of forecasts of $W_{t+1}$. These are defined as:
        \begin{subequations}
        \begin{align}
            & S_{a,t}=[R_{a}, B_{a}]_t^T \\
            &  B_{a,t}=[\tilde{\lambda}_{\text{buy/sell}}, \tilde{\gamma}_{\textrm{EV}}, \tilde{P}_{\text{PV}}, \tilde{P}_{\text{ST}}, \tilde{P}_{\text{load}}^{\text{e}}, \tilde{P}_{\text{load}}^{\text{th}}]_{t}^T
        \end{align}
        \end{subequations}
    } 
    \item{The superscripts $\textrm{e}$ and $\textrm{th}$ refer to electricity or thermal carriers. They are used when the subscript is the same.}
    \item{Both the actions and state vectors have upper and lower limits denoted as $\overline{P}_{a,t}^*$, $\underline{P}_{a,t}^*$, $\overline{S}_{a,t}$, and $\underline{S}_{a,t}$.}
    \item{All bidirectional powers, either actions or states, are modeled with their conversion efficiency $\eta_a$:
        \begin{equation}
       \eta_a S^+_{a,t} - \frac{1}{\eta_a} S^-_{a,t} = S_{a,t} \,,
        \end{equation}
        \noindent{with $S_t^- \perp S_t^+$}
        }
    \item {The order of the subscript is \textit{"name, device, time index"}.}
    \item {Capital $C$ denotes total cumulative cost in \texteuro, lowercase $c$ denotes unit cost and lowercase $w$ indicates tunning/scaling weight.}
\end{itemize}

In Eq. \ref{eq:sdp} the planner or policy $\pi$ wants to minimize the likelihood of the operational cost $\mathbb{E}[C]$ under the exogenous information process $W$. The problem at hand is a state-dependent problem in which our decisions $P_{a,t}^*$ are based on the current $S_{a,t}$, and influence future states $S_{a,t+1}$ (and thus, future decisions).  Given the focus on future states and decisions, lookahead policies appear as attractive candidates for solving this \ac{sdp}. Policy design and models are presented in the following Section \ref{sec:policyDLA}.

\section{Policy design}\label{sec:policyDLA}

\begin{figure}[tb!]
    \centering
    \includegraphics[width=0.9\columnwidth]{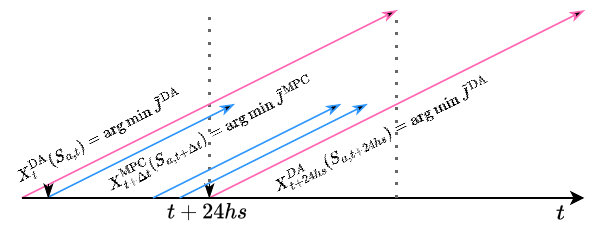}
    \caption{Deterministic DLA policy with a day-ahead planner and economic MPC layer.}
    \label{fig:appDLA}
\end{figure}

As mentioned before,  the \ac{sdp} in Eq. \ref{eq:sdp} is a state-dependent problem where current states influence future decisions. As such, \ac{dla} policies are commonly used in the literature to solve these problems. Two common examples of this policy family are optimal control strategies and stochastic dual dynamic programming. For this work, we focus on economic nonlinear MPC \cite{Grune2017NonlinearAlgorithms}, which is a subset of optimal control where deterministic inputs (forecast medians in this case) are used to decide the actions for the incoming day. The process is shown in Fig. \ref{fig:appDLA}. The policy $X^{\pi}$ takes inputs $\tilde{B}_{a,tt'}$ to decide the power dispatch $\tilde{P}^*_{a,tt'}$ , optimizing over policy time $t'\in [t, t+H^{\text{DA/MPC}}]$ with $H^{\text{DA/MPC}}$ the optimization horizons.

More specifically, a \ac{dla} policy based on the devices' approximated dynamic models $\tilde{S}_t^M(\cdot)$. In this way, the \ac{ems} plans future actions based on approximate predictive models of devices. Approximation is denoted with  $\tilde{}$ . The policy is then solving the implicit \ac{empc} given by:

\begin{subequations}\label{eq:appOCP}
    \begin{align}
        \min_{\tilde{P}^{\text{DA/MPC}}_{a,tt'} \in \mathcal{P}} \quad &  \tilde{J}^{\text{DA/MPC}} \\
        \textrm{s.t.} \quad & \tilde{S}_{a,tt'+1}^{\text{DA/MPC}}=\tilde{S}^M_{a,t} \left(\tilde{R}_{a,tt'}^{\text{DA/MPC}}, \tilde{P}^{\text{DA/MPC}}_{a,tt'}, \tilde{B}^{\text{DA/MPC}}_{a,tt'}\right) \\
                    & \tilde{S}_{a,tt'}^{\text{DA/MPC}} \in \mathcal{S} & \forall a \in \mathbb{A}
    \end{align}
\end{subequations}

\noindent{where:}

\begin{subequations}
\begin{align}
     & \tilde{J}^{\text{DA/MPC}} = \tilde{C}_{\textrm{grid}}^{\text{DA/MPC}}+\tilde{C}_{\textrm{loss}}^{\text{DA/MPC}}+\tilde{p}_{\textrm{SoCDep}}^{\text{DA/MPC}}+\tilde{p}_{\textrm{TESS}}^{\text{DA/MPC}}\\
     & \tilde{p}_{\textrm{TESS}}^{\text{DA/MPC}}=w_{\textrm{TESS}}. \sum_{t'=t}^{H} \max\left(0, \tilde{SoC}^{\text{DA/MPC}}_{\textrm{TESS},tt'} - \overline{SoC}_{\textrm{TESS}} \right)\Delta t
\end{align}
\end{subequations}

In our policy, the upper limit constraint of the TESS, $\overline{SoC}_{\textrm{TESS}}$, is implemented as a soft constraint to avoid infeasibilities during initialization or feedback. The penalty in the objective steers the $SoC_{\textrm{TESS},t}$ towards the feasible region when the weight $w_{\textrm{TESS}}$ is high enough. The \ac{empc} is an adaptation of the policy presented in \cite{Slaifstein2026SequentialMPC}. The only difference between the two layers (DA and MPC) is the length of the horizon and the terminal conditions, which will be analyzed in the Section \ref{sse:bess}.

The deterministic optimization problem in Eq. \ref{eq:appOCP} approximates the real stochastic one by using forecasts, stored in $\tilde{B}_{a,tt'}$, and approximated models for the transition function $\tilde{S}_{a,t}^M$. In this model, the time $t$ is the time at which the \ac{dla} policy is created and $t'$ is the time inside the policy itself.  Note the subtle difference between the approximated dynamics $\tilde{S}_{a,t}^M$ and the real ones $S_{a,t}^M$. This is not to be overlooked because the assumption that the predictions made by the policy $\pi$ hold true can lead to disappointing results in real-world applications. Making these distinctions early in design reveals important insights for future stages.  In this work, the \ac{ema} has an approximated model $\tilde{S}_{a,t}^M$ to decide the setpoints $\tilde{P}^*_{a,tt'}$ to be implemented in a simulator $S_{a,t}^M$ containing detailed fidelity models. In the future, the simulator might as well grow enough to be considered a digital twin of the real building.

Thus the policy is:
\begin{equation}
   X^{\textrm{DA/MPC}}_t(S_{a,t})= \arg \min_{P_{a,t}^{\textrm{DA/MPC}}}  \tilde{J}^{\textrm{DA/MPC}}
\end{equation}
\noindent{subject to the approximate transition function $\tilde{S}_{a,t}^M$. This encompasses model approximation and forecasting $(\tilde{B}_{a,tt'}$) of the future inputs ($W_{t+1}$). The policy is then tuned by changing the weights $w$ and implementing different NLP solver options (warm-starting, multi-start, etc.)}

The approximate transition function $\tilde{S}_{a,t}^M(.)$ is the compendium of the equations specified in the rest of this section. In the remainder of this section, all equations will be presented just in terms of $t$ for the sake of simplicity. However, the reader must remember that when inside the policy $X^{\pi}$ they are defined under the policy's time $t'$.

\subsection{Power \& thermal balances}
Assuming an indoor temperature setpoint decided either by the user or the thermal system \cite{Alpizar-Castillo2024ModellingHouse, Damianakis2023Risk-averseConsumption, Slaifstein2026SequentialMPC},  the building's thermal balance comes in as:
\begin{equation}\label{eq:thBalance}        \tilde{P}_{\text{ST},t}+P_{\text{HP},t}^{\textrm{th}}+P_{\text{TESS},t}=\tilde{P}_{\text{load},t}^{\textrm{th}} \,.
\end{equation}
The electric power balance, on the other hand, is:
\begin{equation}\label{eq:powerBalance}
    \tilde{P}_{\text{PV},t}+P_{\text{BESS},t}+ \gamma_{\text{EV},t}.P_{\text{EV},t}+P_{\text{grid},t}=\tilde{P}_{\text{load},t}^{\textrm{e}}+P_{\text{HP},t}^{\textrm{e}} \,.
\end{equation}
\noindent{where $\gamma_{\textrm{EV}}$ is the \ac{ev} availability, explained in Section \ref{sse:ev}.}

\subsection{Thermal modelling}
The thermal assets are modelled in a linear way. The building has an air-to-water heat-pump that transforms electric power $P_{\text{HP},t}^{\text{e}}$  to heat-flow $P_{\text{HP},t}^{\text{th}}$ with a coefficient of performance $\eta_{\text{HP}}$:

\begin{equation}
    P_{\text{HP},t}^{\textrm{th}}=\eta_{\text{HP}}.P_{\text{HP},t}^{\textrm{e}} \,,
\end{equation}
\noindent{where $\eta_{\text{HP}}$ is assumed constant due to low variation of it during the year \cite{Alpizar-Castillo2024ModellingHouse}.}

The thermal solar collector converts irradiance into heat, and assuming a constant linear relationship with the generated \ac{spv} power $P_{\text{PV},t}$ as:
\begin{equation}
    \tilde{P}_{\text{ST},t}=\eta_{\text{ST}}.\tilde{P}_{\text{PV},t} \,,
\end{equation}

This work considers an underground, perfectly mixed water tank with independent charge and discharge coils as a \ac{tess}. Assuming no mass exchange between the \ac{tess} and the piping system the \ac{tess} the state of charge is:
\begin{equation}\label{eq:socTESS}
SoC_{\text{TESS},t+1}=SoC_{\text{TESS},t}-\frac{\Delta t}{Q_{\text{TESS}}.3600}.\eta_{\text{TESS}}.P_{\text{TESS},t} \,,
\end{equation}
\noindent{where $\eta_{\text{TESS}}$ denotes the heat-exchanger efficiency, $Q_{\textrm{TESS}}$ is the capacity in kWh. To avoid imposing an arbitrary periodicity on the \ac{tess}, no terminal conditions are used on its $SoC_{\text{TESS},t}$. Thus, the \ac{tess} operates under its natural time constant.}

\subsection{Batteries}\label{sse:bess}
\begin{figure}
    \centering
    \includegraphics[width = 0.5\columnwidth]{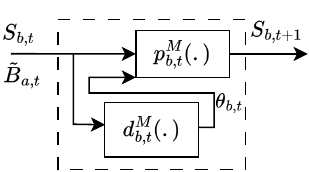}
    \caption{Battery storage asset transition function diagram $\tilde{S}^M_{b,t}$}
    \label{fig:saModels}
\end{figure}

The remaining devices in the \ac{mces} are all battery-based \ac{ess}. Batteries have complex nonlinear dynamics, and several modeling techniques are presented in the literature \cite{Plett2015}. In this work, models coming from empirical and physics-based approaches are used. The modeling is divided into two different sub-models: performance and ageing. Under the \ac{umf}, this is represented in the transition function $\tilde{S}^M_{b,t}(\tilde{S}_{b,t},x_{b,t}|\theta_{b,t})$, which contains both the perf. model $p_{b,t}^M(.)$ and the ageing model $d_{b,t}^M(.)$. The performance model predicts stored energy $SoC_{b,t}$ and terminal voltage $v_{t,b, t}$.  The ageing model is used to update the parameters of $p^M_{b,t}(.)$, as shown in Fig. \ref{fig:saModels}. Even though the change in parameters $\theta_{b,t}$ becomes significant after considerable ageing has occurred, optimizing it in the short term can lead to considerable savings in the long and medium term \cite{Cao2020MultiscaleModels, Movahedi2024ExtraMechanism, Jin2022, Reniers2021}. This is because in \ac{pb} ageing models, the relationship between ageing states and control actions is explicit, and the policy can directly minimize it through the internal states.

The perf. model is then:
\begin{equation}
S_{b,t+1}=p^M_{b,t}\left(S_{b,t}, P_{b,t}, B_{a,t}|\theta_{b,t}\right)
\end{equation}
\noindent{where the components of the state depend on the functional form used for the model. In general, this is a nonlinear state space system.}

The ageing model $d_{b,t}^M(.)$ is a set of equations that describes the dynamics of the performance parameters $\theta_{b,t}$.
\begin{equation}
\theta_{b,t+1}=d_{b,t}^M\left(S_{b,t}, P_{b,t}, B_{a,t}, \theta_{b,t}\right)
\end{equation}

Finally, assuming that there is a strong daily usage pattern a terminal constraint is implemented to ease up feasibility and mitigate symmetries in the \ac{ocp} of Eq. \ref{eq:appOCP}. Following \cite{Slaifstein2026SequentialMPC}, the daily periodical constraint in the first optimization of the day ($X^{\text{DA}}_t$ in Fig. \ref{fig:appDLA}) is:
\begin{equation}
  \tilde{SoC}_{\textrm{BESS},tt'_1}^{\textrm{DA}} = \tilde{SoC}_{\textrm{BESS},tt'_1+24hs}^{\textrm{DA}} \label{eq:pcDA} 
\end{equation}
\noindent{and in all other steps $X^{\text{MPC}}_t$ the periodic condition is:}
\begin{equation} \label{eq:pcCT}
    \tilde{SoC}_{\textrm{BESS},tt'_0}^{\textrm{MPC}} = \tilde{SoC}_{\textrm{BESS},tt'_0+H^{\textrm{MPC}}}^{\textrm{MPC}}
\end{equation}
\noindent{where $t_0 \leq t_1 \leq H \leq T$. This way, the \ac{ocp} is better computationally conditioned, but the planner still has the freedom to decide the state-of-charge at the begining of each day ($SoC_{\textrm{BESS},24\textrm{hs}}$). The proposed terminal condition has two key properties: it is more flexible than fixing $SoC_{\text{BESS},24\text{hs}}=50\%$ , and it bounds the value function $V_H$ of the \ac{ocp}. Ideally, no terminal condition would be used to freely use all 3 storage systems. Unfortunately, to solve such an unbounded \ac{ocp}, an optimization horizon $H$ much larger than 48 hours would be required \cite{Grune2017NonlinearAlgorithms, Prat2024HowProblems}.}

\subsubsection{Performance models $p_{b,t}^M$}
For the performance submodel, two alternatives have been implemented: a simple Coulomb counting or \ac{bm} and a first-order \ac{ecm}. A basic \ac{bm} of the operation of a battery assumes that its output voltage $v_t$ is linear with the state of charge $SoC$, assuming no internal resistance, and no-diffusion dynamics. Hence, the only equations of this model are:
\begin{subequations}\label{eq:BM}
\begin{align}
    \label{eq:soc}
    & SoC_{b,t+1}=SoC_{b,t}-\frac{\Delta t}{Q_{b,t}.3600}.\eta_c.i_{b,t} \,,\\
    \label{eq:ibranch}
    & i_{b,t}=\frac{P_{b,t}}{v_{t,b,t}.N_{s,b}.N_{p,b}} \,,\\
    \label{eq:ocv}
    & OCV_{b,t}=a_{OCV,b}+b_{OCV,b}.SoC_{b,t} \,,\\
    \label{eq:vtBM}
    & v_{t,b,t}=OCV_{b,t} \,,\\
    & S_{b,t}=[SoC_{b}, v_{t,b}, i_{b}]^T_t
\end{align}
\end{subequations}
\noindent{where $i_{b,t}$ is the current passing through the cell, $OCV_{b,t}$ is the open circuit voltage, $\eta_c$ is the Coulombic efficiency \cite{Plett2016} , and $Q_{b,t}$ is the cell capacity in Ah. Each battery pack is assumed to be organized as a \ac{scm} where  $N_{s/p,\ b}$ are the series cells per branch and parallel branches, respectively. In this model, the most relevant parameter in $\theta_{b}$ is the $Q_{b}$.} 

\begin{figure}
\centering
\includegraphics[width=0.5\linewidth]{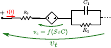}
	\caption{First-order Equivalent Circuit Model.}
	\label{fig:ecm}
\end{figure}

A first-order \ac{ecm} has improved accuracy due to the incorporation of diffusion and series resistance, as in Fig. \ref{fig:ecm}. The performance sub-model $p_{b,t}^M(.)$ is then modified by adding the equation:
\begin{equation}\label{eq:ecmRC}
    i_{R_1,b,t+1}=e^{-\frac{\Delta t}{R_{1,b}.C_{1,b}}}.i_{R_1,b,t}+ \left( 1-e^{-\frac{\Delta t}{R_{1,b}.C_{1,b}}} \right) .i_{b,t}
\end{equation}
\noindent{and modifying Eq. \ref{eq:vtBM} as in:}
\begin{equation}\label{eq:vtECM}
    v_{t,b,t}=OCV_{b,t}-i_{R_1,b,t}.R_{1,b}-i_{b,t}.R_{0,b} \,,
\end{equation}
\noindent{where $i_{R_1,b,t}$ is the current flowing through $R_1$ in Fig. \ref{fig:ecm}. Eqs. \ref{eq:soc}, \ref{eq:ibranch} and \ref{eq:ocv} are maintained. The \ac{ecm} incorporates the series voltage drop that limits power output and the first-order diffusion dynamics. Here, the relevant parameters are $\theta = [Q, R_0]^T$ that usually define the cell's state of health $SoH$.}

\subsubsection{Degradation models $d_{b,t}^M(.)$}
\label{sse:degModel}
For the ageing models, the first alternative is an empirical sub-model presented by \cite{Wang2014}. The empirical sub-model reduces all the degradation mechanisms into calendar and cyclic ageing.
\begin{subequations}\label{eq:empDeg}
    \begin{align}
        \label{eq:cycliclLoss}
        & i_{\text{cycle},b,t}=\frac{c_1.c_3}{c_4}.e^{c_2.|i_{b,t}|}.(1-SoC_{b,t}).|i_{b,t}| \,, \\
        \label{eq:calLoss}
        & i_{\text{cal},b,t}=c_5.e^{-\frac{\text{24 kJ}}{RT}}.\sqrt{t_{0,b}+t} \,, \\
        & i_{\text{loss},b,t}=i_{\text{cycle},b,t}+i_{\text{cal},b,t} \,,
    \end{align}
\end{subequations}
\noindent{and}
\begin{equation}\label{eq:Qdyn}
Q_{b,t+1}=Q_{b,t} -\frac{\Delta t}{3600}.i_{\text{loss},b,t} \,.
\end{equation}

\noindent{where the fitting parameters $c_{1,...,5}$ are taken from \cite{SlaifsteinIECON23, Vermeer2022b, Wang2014} and $t_{0,b}$ is the elapsed lifetime of the battery $b$.}

For the physics-based alternative, the reduced order model (\ac{pbrom}) from \cite{Jin2022}  is used. It accounts for two degradation mechanisms: the \ac{sei} and \ac{am}. The author also presents a \ac{pbrom} for Li-plating, but given the low C-rate and standard temperature range of this application, it will not be included. 

The growth of the \ac{sei} layer is modeled with a general reaction that aims to average all the different byproducts that compose the \ac{sei} layer. This is synthesized in the reversible \ac{sei} current $i_{\textrm{SEI}}$:
\begin{equation}\label{eq:iSEI}
    i_{\textrm{SEI},b,t} = \frac{k_{\textrm{SEI},b}.e^{\frac{-E_{\textrm{SEI},b}}{RT}}}{n_{\textrm{SEI}}.(1+\lambda_{b}.\beta_{b}).\sqrt{t_{0,b}+t}}
\end{equation}
\noindent{where $k_{\textrm{SEI},b}$ is the kinetic rate of the average reaction, $E_{\textrm{SEI},b}$ is the activation energy of the reaction, $n_{\textrm{SEI}}$ is the average number of $e^-$ transferred through the layer, and $\lambda_b$ and $\beta_b$ are parameters depending on other variables such as $\eta_{k,b}$, $OCV_{n,b}$, $z_b$ and others.}

The system is completed with:
\begin{subequations}
\begin{align}
    & \eta_{k,b,t}=\frac{2.R.T}{F}.\text{sinh}^{-1}\left(\frac{i_{b,t}}{n_{\textrm{SEI}}.a_s.A.L_n.i_0}\right) \\
    & z_{b,t} = SoC_{b,t}.(z_{100\%,b}-z_{0\%})+z_{0\%,b} \\
    & \beta_{b} = e^{\frac{n_{\textrm{SEI}}.F}{R.T}.\left(\eta_{k,b}+OCV_{n,b,t}-OCV_{s,b} \right)}
\end{align}
\end{subequations}

\noindent{where $\eta_{k}$ is the SEI side reaction kinetic overpotential, $z$ is the Li stochiometry of the cell, $OCV_{n}$ is the open-circuit voltage of the anode made with an empirical fit, $OCV_s=0.4V$ is the side reaction open-circuit voltage, and $T$ is the cell temperature. It is assumed that the temperature $T$ is constant over time and is controlled by the local primary control system of the \ac{bess} and \ac{ev}. The rest of the parameters can be found in the  \ref{sec:appA}.}

The loss of active material due to the mechanical stress of the electrode is modeled with:
\begin{equation}
        i_{\textrm{AM},b,t} = k_{\textrm{AM},b}.e^{\frac{-E_{\textrm{AM},b}}{R.T}}.SoC_{b,t}.|i_{b,t}|.Q_{b,0}
\end{equation}
The total ageing is the contribution of both mechanisms \ac{sei} layer growth and \ac{am} loss. The capacity fade current is:
\begin{equation}\label{eq:pbDeg}
    i_{\text{loss},b,t} = i_{\textrm{SEI},b,t} + i_{\textrm{AM},b,t} 
\end{equation}
\noindent{which is later used again in \ref{eq:Qdyn}.}

Now, by carefully inspecting Eq. \ref{eq:vtECM}, the reader will notice that if $R_{0,b,t}$ is incorporated as a variable in the OCP, Eq. \ref{eq:appOCP}, this would add another non-convex constraint to it (since $i_{b,t}$ can be either positive or negative). Thus, its evolution is only included in the simulator $S^M_{a,t}(.)$ updating the parameters without the policy $X^{\pi}_t$ being directly aware of the process.

To model the power fade (i.e., the increase of $R_0$), the \ac{sei} layer thickness $\delta_{\textrm{SEI},b,t}$ growth is described by:
\begin{equation}
    \delta_{\textrm{SEI},b,t+1} = \delta_{\textrm{SEI},b,t} + \frac{\Delta t}{M_{\textrm{SEI}}.n_{\textrm{SEI}}.F.\rho_{\textrm{SEI}}.A_n} i_{\textrm{SEI},b,t}
\end{equation}
Hence, the dynamics of the series resistance $R_0$ are:
\begin{equation}
    R_{0,b,t+1} = R_{0,b,t} + \frac{\varepsilon_s}{\kappa_{eff}}.\frac{\Delta t}{M_{\textrm{SEI}}.n_{\textrm{SEI}}.F.\rho_{\textrm{SEI}}.A_n}i_{\textrm{SEI},b,t}
\end{equation}
The solvent S leaves the electrolyte to form the SEI layer; thus, the volume fraction of S evolves with:
\begin{equation}
    \varepsilon_{e,b,t+1} = \varepsilon_{e,b,t} - a_s. \frac{\Delta t}{M_{\textrm{SEI}}.n_{\textrm{SEI}}.F.\rho_{\textrm{SEI}}.A_n}i_{\textrm{SEI},\ b,\ t}
\end{equation}

\subsection{Electric Vehicle}\label{sse:ev}

From the point of view of a residential building, the \ac{ev}s are a \ac{bess} with availability constraints and certain requirements regarding their $SoC$ at departure time $t_{\text{dep}}$. For the availability $\gamma$, the probability distributions of departure ($t_{\text{dep}}$) and arrival ($t_{\text{arr}}$) times can be described as random variables $t_{\textrm{dep/arr}} \sim \mathcal{T_{\textrm{dep/arr}}}$, whose distributions $\mathcal{T}_{\text{dep/arr}}$ are taken from Elaad \cite{elaad}. The availability $\gamma_t$ will then be:
\begin{equation}
\gamma_t=\begin{cases}
      0 & t \in [t_{\text{dep}};\ t_{\text{arr}}]\\
      1 & \text{otherwise}
    \end{cases} \,.
\end{equation}
The power balance of an \ac{ev} is
\begin{equation}
P_{\textrm{tot},\textrm{EV},t}=\gamma_{\textrm{EV},t}.P_{\textrm{EV},t}+(1-\gamma_{\textrm{EV},t})P_{\textrm{drive},\textrm{EV}}
\end{equation}
\noindent{where $P_{\text{tot}, \text{EV}, t}$ is the total power of the \ac{ev}, $P_{\text{EV}, t}$ is the charger power, and $P_{\text{drive},\text{EV}}$ is the power consumed driving, assuming no public charging. The total power $P_{\text{tot}, \text{EV}, t}$ is then used in Eq. \eqref{eq:ibranch} and later for calculating the ageing of the \ac{ev} batteries. The average driving power is also sampled from a Gaussian distribution $P_{\text{drive},\text{EV}} \sim \mathcal{N}(\mu_{\text{drive}},\,\sigma_{\text{drive}}^{2})$. This is because the \ac{ev} battery pack degradation during driving needs to be accounted for in the operation strategy (charging and driving).}

At $t_{\textrm{dep}}$ the \ac{ev} is required to be delivered at $SoC_{\textrm{dep}}^*$:
\begin{equation}
SoC_{\textrm{EV}}(t_{\textrm{dep}})=SoC_{\textrm{dep}}^*
\end{equation}
This is implemented as a penalty in the objective function, Eq. \ref{eq:appOCP}, as in any typical OCP. The deviation from the reference at the desired time is penalized with:
\begin{equation}
\varepsilon_{\textrm{SoC}, t_{\text{dep}}}=SoC_{\textrm{EV}}(t_{\textrm{dep}})-SoC_{\textrm{dep}}^*
\end{equation}

Summing up, the three different storage systems have three different ways of handling their terminal set and value function bounds. The \ac{bess} has a periodic conditions Eqs. \ref{eq:pcDA} and \ref{eq:pcCT} , which is a flexible alternative to the classic terminal daily periodicity. The \ac{ev} follows a soft-tracking problem at $t_{\text{dep}}$. The \ac{tess} has no terminal condition. This flexible design is the key contribution to integrating all the storage systems and the non-convex \ac{pb} models to the \ac{empc} of Eq. \ref{eq:appOCP}.

\begin{algorithm}[!t]
    \caption{\ac{mces} simulation}\label{alg:Sim}
    \begin{algorithmic}[1]
	\State \textbf{Define setpoint} $P_{a,t}$ \;
        \State \textbf{Define exogenous information} $W_{t+1}$ \;
        \State Recalculate $P_{\textrm{TESS},t}$ and $P_{\textrm{grid},t}$ using Eqs. \ref{eq:thBalance} \&  \ref{eq:powerBalance}, with $P_{a,t}$,  and $W_{t+1}$ \;
        \State Simulate $SoC_{\textrm{TESS}, t}$ using  Eq. \ref{eq:socTESS} \;
        \State Simulate $b$ performance using  $p_{b}^M(.)$ \ac{pbrom} \cite{Planden2022}  \;
        \State Simulate $b$ degradation using  $d_{b,t}^M(.)$ \ac{pbrom} \cite{Jin2022}  \;
        \State Feed-back $S_{a,t}$ to the planner \;
    \end{algorithmic}
\end{algorithm}

\section{MCES Simulator $S_{a,t}^M(.)$}\label{sec:simMCES}

The simulator evaluates the policy $\pi$ and closes the loop with the state measurements. It is designed to:
\begin{itemize}
    \item Provide high-accuracy simulation results that act as plant measurements.
    \item Adjust/reject setpoints that violate hard constraints.
    \item Re-balance power in case of rejections or infeasible optimizations.
\end{itemize}

The whole process is defined in Algorithm \ref{alg:Sim}. First, the power setpoints must be adjusted for the grid and \ac{tess} because the forecast used in $X_t^{\pi}$ will never be the same as the actual exogenous inputs. Take a look at the balances, Eq. \ref{eq:thBalance} and \ref{eq:powerBalance}, which contain the loads and solar generation. It is clear that $\tilde{P} \neq P$ and a device must compensate for that difference.  Thus, the simulator $S_{a,t}^M$ recalculates:
\begin{equation}
P_{\text{TESS},t}=P_{\text{load},t}^{\textrm{th}} - P_{\text{ST},t} -P_{\text{HP},t}^{\textrm{th}} \,.
\end{equation}
\begin{equation}
P_{\text{grid},t}=P_{\text{load},t}^{\textrm{e}}+P_{\text{HP},t}^{\textrm{e}} -P_{\text{PV},t} - P_{\text{BESS},t} - \gamma_{\text{EV},t}.P_{\text{EV},t} \,.
\end{equation}

Second, once these powers have been adjusted, the simulator uses these powers to obtain the true/actual/fidelity state $S_{a,t}$. For the \ac{tess} it recalculates Eq. \ref{eq:socTESS}. For the $b$, it uses LiiBRA.jl \cite{Planden2022} to swiftly simulate \ac{pbrom}s of the performance of the battery \cite{Plett2015, Rodriguez2019}. After that, the models from Jin \cite{Jin2022} are used to calculate the true degradation outcome of the decisions $P_{a,t}^*$. Again, the reader must remember that the capacity fade (decrease in $Q_{b,t}$) is modeled in both the simulator $S^M_{a,t}$ and the approximate model of the planner $\tilde{S}_{a,t}^M$, whereas the power fade (increase in $R_{0,b,t}$) is only addressed in the simulator $S^M_{a,t}$. Finally, if an action $P_{b,t}^*$ causes a future state to go out of bounds ($S_{b,t+1}\leq \underline{S}_{b}$ or $S_{b,t+1}\geq \overline{S}_{b}$), the action is rejected and modified leaving $b$ on the bound (either $\overline{S}_{b}$ or $\underline{S}_{b}$) until the next $t$. Finally, the carriers are rebalanced if necessary.

The final state $S_{a,t}$ is then fed back to the optimization-based controller. For practical implementation, in which the simulator is, in fact, an experimental setup, an online state observer is necessary to feed back the states to the \ac{ems}. This is particularly important for the \ac{ess} \cite{Plett2015, Plett2016, Plett2024BatteryMethods, Fan2023NondestructivePaths}.

\begin{figure}[bt!]
    \centering
    \begin{subfigure}{0.55\textwidth}
        \includegraphics[width=0.9\linewidth]{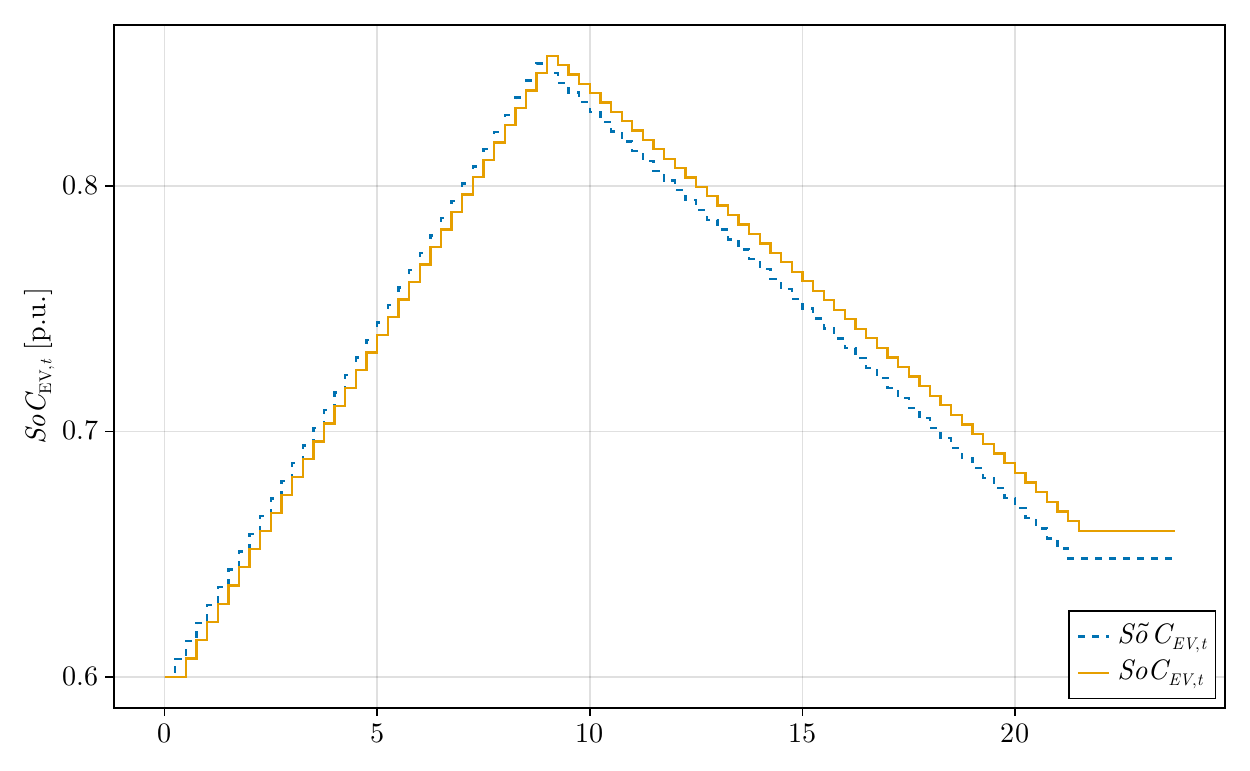} 
        \caption{Simulated and predicted $SoC$ for the \ac{ev} for $T=24$hs and open-loop simulation.}
        \label{fig:ctrl_rlt_EV}
    \end{subfigure}
    \begin{subfigure}{0.44\textwidth}
        \includegraphics[width=0.9\linewidth]{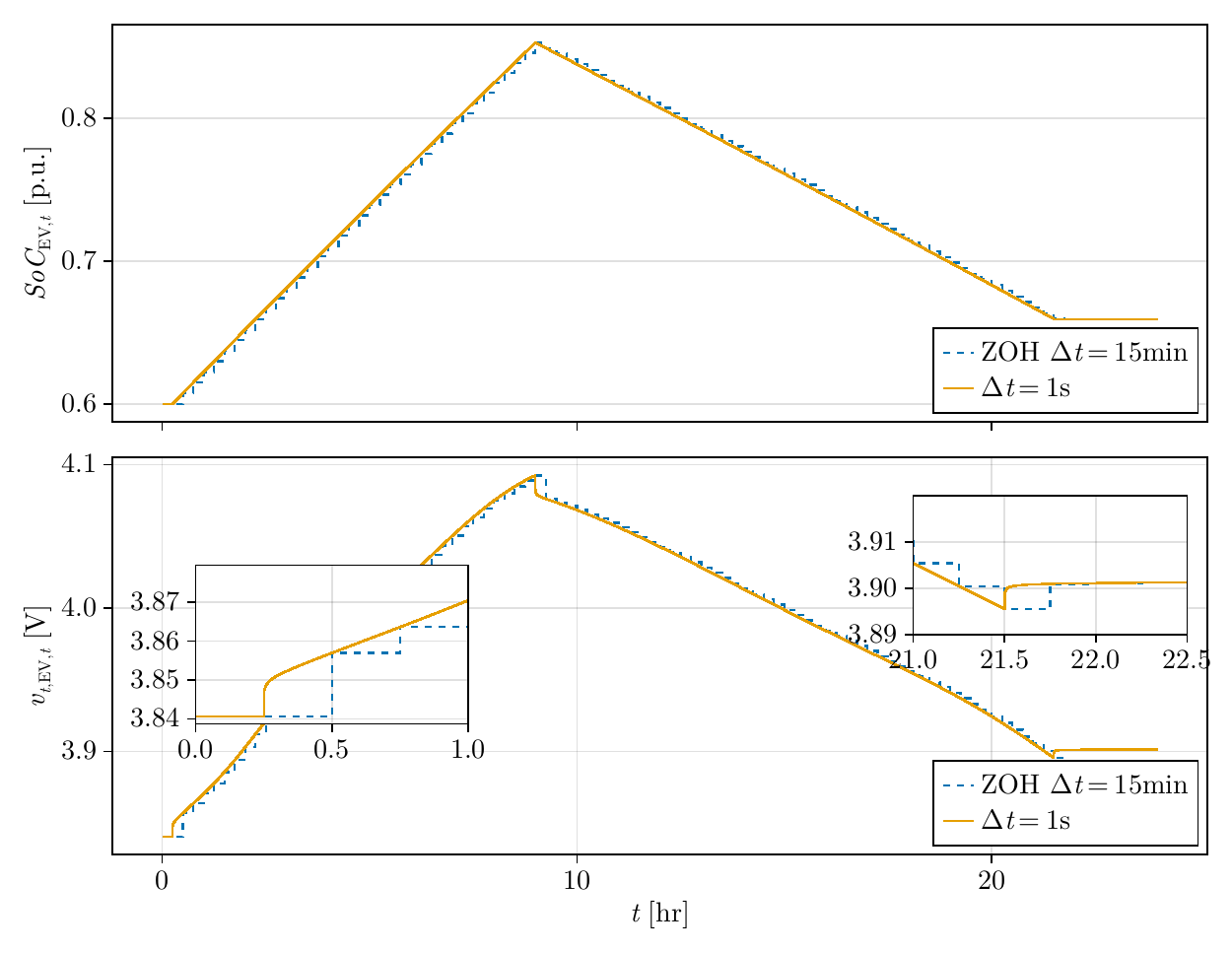}
        \caption{Full simulation and reported down sampled state.}
        \label{fig:fullRes}
    \end{subfigure}
    \caption{}
    \label{fig:sim_pred_ev}
\end{figure}

Despite the inherent inaccuracies of the simplified models used within the \ac{ems} planners $\tilde{S}_{a,t}^M(\cdot)$, the informed $SoH_{b,t}$ remains accurate due to the high-fidelity feedback provided by $S_{b,t}^M(\cdot)$. It is assumed that the primary control of the storage $b$ will maintain the current spikes within tight bounds, avoiding any consequent voltage deviations within the sampling time $\Delta t =15$min of the \ac{ems}.  Thus, since the simulator $S_{b,t}^M(\cdot)$ operates at a $\Delta t_s=1$s resolution, the informed degradation captures the true physical response of the battery, including peak state values. These are later down-sampled to the $\Delta t=15$min resolution of the planner $X^{\pi}$. The process effectively reduces the planner's error to a mere 15-minute communication lag (Fig. \ref{fig:sim_pred_ev}) and a zero-order-hold approximation. By utilizing the realized trajectories of $SoC_{b,t}$ and $v_{t,b,t}$ from the fidelity simulator $S_{a,t}^M(\cdot)$ rather than the planner's internal estimates, the system ensures that the reported capacity fade and degradation states reflect the actual operational stress on the battery, regardless of the planner's complexity.

\begin{figure}[bt]
    \centering
    \includegraphics[width=\columnwidth]{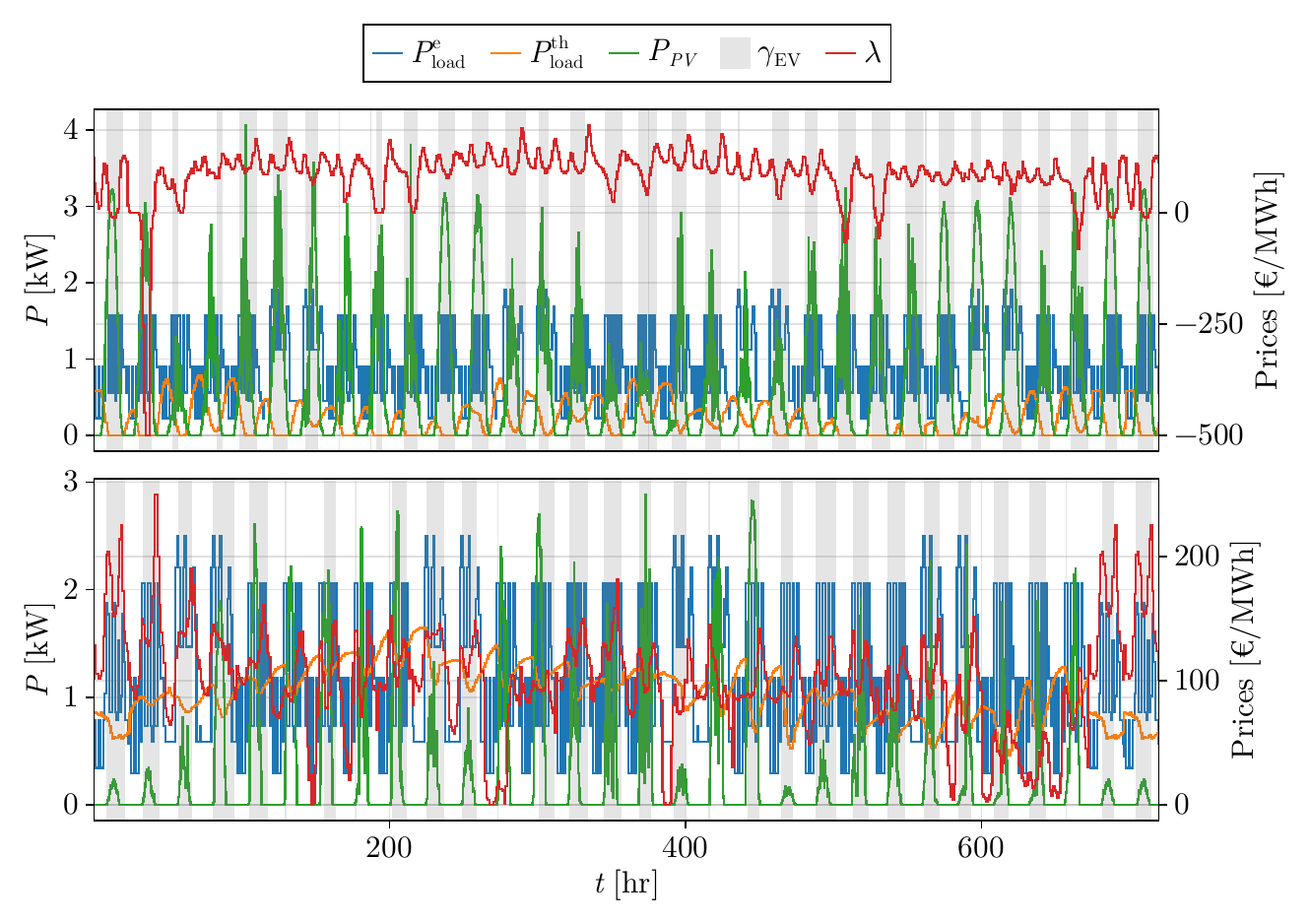}
    \caption{Exogenous information $W_{t+1}$. Grey bands represent periods were the \ac{ev} is not connected.}
    \label{fig:cs1_inputs}
\end{figure}

\section{Case studies}
\label{sec:results}

\begin{algorithm}[b!]
    \caption{Rolling horizon algorithm}\label{alg:CS1-RH}
    \begin{algorithmic}[1]
	\State \textbf{Initialize hyperparameters} $t_0,\ \Delta t,\ H^{\text{DA/MPC}},\ w,\ n_d$ \;
    \State \textbf{Initialize device states and inputs} $S_{a,\ 0}$ \;
    \For{$d \in 1:n_d$}
        \State Solve the day-ahead policy $X^{\text{DA}_t}$ and obtain schedule $\mathcal{P}_{a,[t,H^{\text{DA}}]}^{\textrm{DA}}$. \;
        \For{$t \in 0:H^{\text{MPC}}$}
        \State Solve the deterministic continuous time policy $X^{\text{MPC}_t}$ and obtain action $P_{a,t}^{\textrm{MPC}}$. 
        \State Simulate $S_{a,t+1}=S_{a,t}^M(S_{a,t}, P^{\textrm{MPC}}_{a,t}, W_{t+1})\ $
        \State Update forecasts in $B_{a,tt'}^{\text{MPC}}$\;
        \State Move time window $t \leftarrow t + \Delta t^{\text{MPC}}$;\;
        \EndFor
    \EndFor
    \end{algorithmic}
\end{algorithm}

The building has a grid connection with a smart meter with 15min resolution. The grid power $P_{\textrm{grid}}$ is included in the state vector $S_{a,t}$. The system is composed of a 4kWp \ac{spv}, a 15.6kWh/12.5kW \ac{bess} with \ac{nmc} or \ac{lfp} cells, one 55.6kWh/12.5kW \ac{ev} charging points, a 4kWe heat pump, a 2.7kWth solar thermal collector, a 200kWh \ac{tess}, a 2.5kWp electrical load, a 1.5kWp thermal demand, and 17kW LV grid connection. Power consumption profiles ($P_{\textrm{load}}^{\textrm{e}}$) were constructed for a year using data from 2021 to 2023 from TU Delft's Green Village smart meter data \cite{GreenVillage}.  The output of the \ac{spv} is taken from \cite{Smets2016, Diab2023}, the market prices $\lambda$ are taken from the EPEX day-ahead auction, and $\lambda_{\textrm{buy}}=0.95\lambda_{\textrm{sell}}$ \cite{2023EPEXServices},  and the heat demand $P_{\textrm{load}}^{\textrm{th}}$ was modeled as \cite{Alpizar-Castillo2024ModellingHouse}. 

The cells used are SANYO NCR18650 cells for \ac{nmc} as in \cite{Jin2022} and A123 cells for \ac{lfp} \cite{Prada2013Simulations}. Their datasets were taken from \jlinl{PyBaMM} \cite{Sulzer2021PythonPyBaMM} and \jlinl{LiiBRA} \cite{Planden2022}. The \ac{ecm}s for both cell types were fitted from synthetic cells simulated in \jlinl{PyBaMM}, following the methodology in \cite{Plett2015}. Once the simulated profiles are ready, the \ac{ecm} parameters can be identified using subspace system identification as in \cite{Plett2004b}. The parameters can be found in \ref{sec:appA}, Tables \ref{tab:parametersPBROM} and \ref{tab:parametersECM} along with a more detailed explanation of the identification pipeline. The capacity fade cost is assumed to be $c_{\text{loss}}=1.2$ \texteuro/Ah, roughly 280-310\texteuro/kWh, depending on the average voltage.

The simulations were modelled and run using \jlinl{Julia} \cite{julia}, \jlinl{JuMP} \cite{JuMP}, and \jlinl{InfiniteOpt} \cite{InfiniteOpt}. The chosen solver was KNITRO from Artelys \cite{KNITRO}. All simulations were run using an Intel CPU at 2.60GHz, 4 processors, and 32GB of RAM. The base library for building and simulating \ac{ems} policies is open-source and free to use in \href{https://github.com/DarioSlaifsteinSk/EMSmodule}{\jlinl{EMSmodule}} \cite{Slaifstein2026SequentialMPC}. Since this is a non-linear non-convex \ac{ocp}, global optimality can not be generally guaranteed \cite{Boyd2011ConvexOptimization, Grune2017NonlinearAlgorithms}. Having this in mind, the implementation of the \ac{ems} approaches this limitation by: (i) using warm-starts as initial guess to ensure convergence to the same local optima, (ii) using the KNITRO tuner and multi-start in each individual step to increase the chance of finding the best local optima \cite{KNITRO}.

\subsection{Case Study I: Degradation model comparison}\label{sse:CS1}
To test and validate our \ac{ema} the closed-loop control was simulated for two standard months (summer and winter) using 2023 data from the previously mentioned sources. Both are presented in Fig. \ref{fig:cs1_inputs}. To quantify the impact of each performance and ageing model 3 \ac{empc} controllers, 2 benchmarks and our proposed approach, were implemented:
    \begin{itemize}
        \item  (\textit{BNoDeg}) Including a bucket model and no degradation $w_{\text{loss}}=0$, with $\tilde{S}^{M1}_{a,t}$ in Eq. \ref{eq:appOCP}.
        \item (\textit{CEmpDeg}) Including a first-order \ac{ecm} and empirical ageing for the $b$, with $\tilde{S}^{M2}_{a,t}$ in Eq. \ref{eq:appOCP}.
        \item (\textit{CPBDeg}) Including first-order \ac{ecm} with \ac{pbrom} ageing for the $b$, with $\tilde{S}^{M3}_{a,t}$ in Eq. \ref{eq:appOCP}.
    \end{itemize}
\begin{table}[h]
    \centering
    \begin{tabular}{ccccc}
    \hline
         \textbf{Battery}
& \multicolumn{3}{c}{\textbf{Scheduler}} & \textbf{Simulator} \\
         \textbf{model}& \textbf{BNoDeg} & \textbf{CEmpDeg} & \textbf{CPBDeg} & \textbf{PBROM} \\
    \hline
    \textbf{Performance}
         & Bucket model& ECM& ECM& SPM\\
 & Eq. 21& Eq. 21-23& Eq. 21-23&PBROM \cite{Planden2022}\\
    \textbf{Ageing}
         & -
         & Empirical& PBROM & PBROM \\
 & &  Eq. 24-25 \cite{Wang2014}& Eq. 25-29 \cite{Jin2022}&Eq. 25-29 \cite{Jin2022}\\
    \hline
    \textbf{Feature}& & & & \\
    \hline
    \textbf{Dynamic}& Only& ECM \cmark& \cmark& \cmark\\
 \textbf{Profile}& performance.& Emp. Deg \xmark& &\\
    \textbf{Cathode}& No distinction
         & Only NMC& \cmark& \cmark\\
 \textbf{Chemistry}& & & &\\
    \textbf{Used and}& Used cells are& \cmark& \cmark& \cmark\\
 \textbf{New cells}& only smaller cells.& & &\\
    \hline
    \end{tabular}
    \caption{Summary of battery models in each planner.}
    \label{tab:batteryModelSummary}
\end{table}

A summary of the key Eqs. and features of each of the planners can be found in Tabel \ref{tab:batteryModelSummary}. The simulation workflow is presented in Algorithm \ref{alg:CS1-RH} and depicted in Fig. \ref{fig:appDLA}. First, the hyperparameters are initialized. This includes the time window to be optimized $\mathcal{D}_t^{\pi}=[t;t+H]$, the number of days $n_d$, user preferences, the initial state $S_{a,0}$, and weights $w$. In our case $n_d=29$, the weights are $w_{\text{grid}}=1,\ w_{\text{loss}}=0.01$ and $w_{\text{SoCDep}}=w_{\text{TESS}}=1000$, since the first two represent real economic costs and the second represent penalties. At timestep $t$ ,  the \ac{ocp} in Eq. \ref{eq:appOCP}, is solved, obtaining the action $P_{a,t}^*$. Together with the exogenous information $W_{t+1}$ the actions are passed to the simulator $S_{a,t}^M$ to get the feedback state $S_{a,t}$.

\begin{figure*}[hb!]
    \centering
    \includegraphics[width=1.15\linewidth]{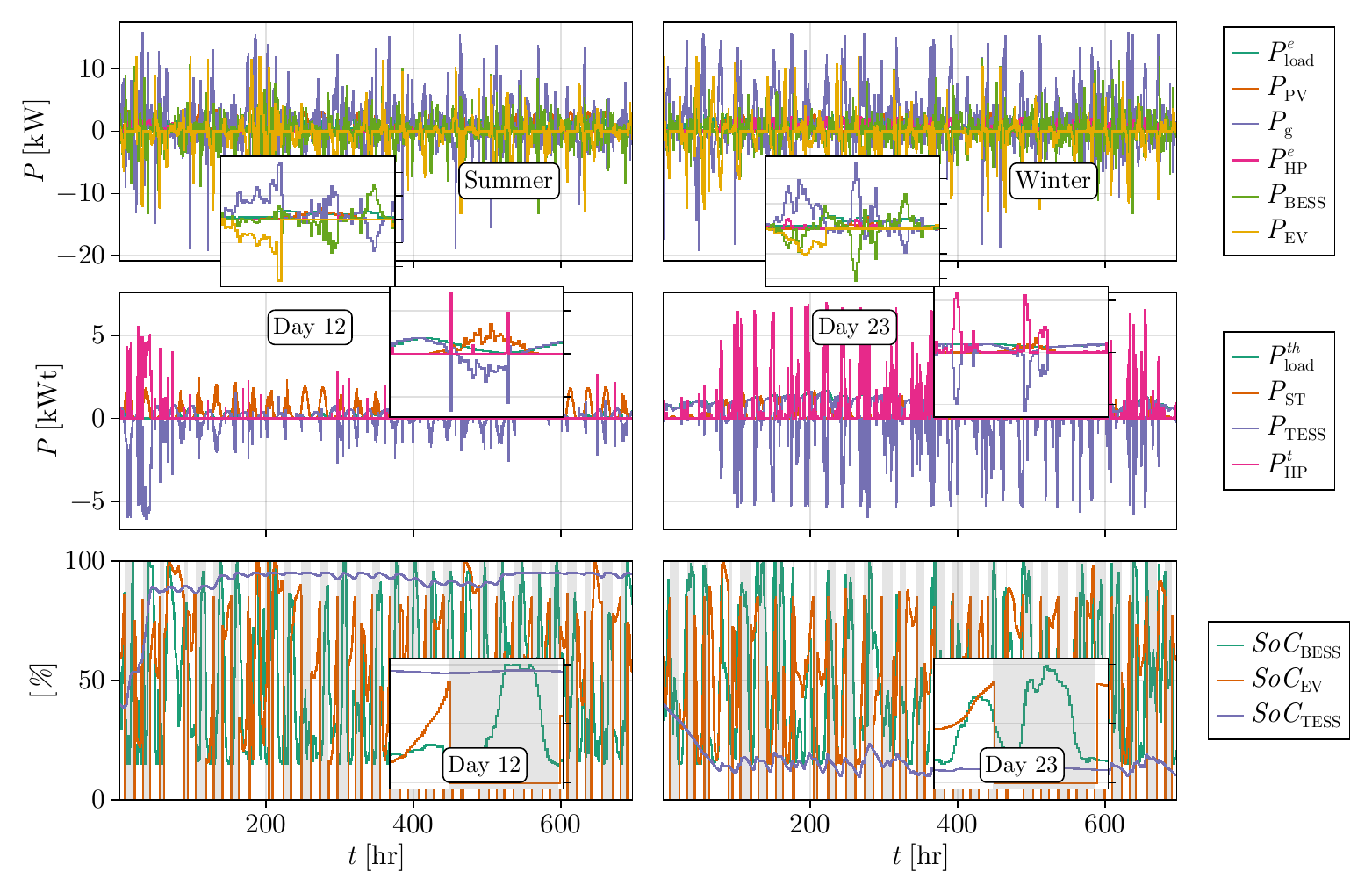}
    \caption{Monthly dispatch of \ac{mces} under \textit{CPBDeg} controller with $w_{\text{loss}}=0.01$ for summer (left) and winter (right).}
    \label{fig:cs1_dispatch}
\end{figure*}

As a representative example, Fig. \ref{fig:cs1_dispatch} presents the results for the proposed \textit{CPBDeg} controller for a monthly period. It has the resulting power balances (electrical and thermal) and the use of the \ac{hess}. The electric \ac{ess} have daily cycles to minimize operating costs (energy arbitrage). This is particularly important for the \ac{ev} since its mobility demand already establishes a daily periodicity. Thus, due to the \ac{ev}'s battery pack size and its natural periodicity, it becomes the main electric storage of the system.  This frees up the \ac{bess} for energy arbitrage, trying to capture price variations when possible within the power balance. Thus, price volatility incentivizes cycling. However, due to the \ac{sei} model, batteries are also pushed downwards to the minimum ageing state at $\underline{SoC}_{b,t}$.  Hence, the dispatch contemplates a trade-off between the 2 parts of the objective $C_{\textrm{grid}}$ and $Q_{\textrm{loss}}$. On the thermal carrier, the $P_{\text{load},t}^{\text{th}}$ represents the heat losses of the building which have to be compensated by either \ac{st}, \ac{hp} or \ac{tess}. In the summer, the losses during solar hours are almost zero and thus the \ac{st} charges the \ac{tess}. If the prices are negative enough then \ac{hp} charges the \ac{tess} during $\lambda_t \leq 0$. In the winter, the \ac{tess} is charged with the \ac{hp} early in the day when prices are low or during solar hours, in that order of priority, to later deliver heat to the demand at high price hours. In summary, during the summer price volatility is high with several hours with $\lambda_t \leq 0$, the \ac{ems} buys this energy to reduce costs. In winter, prices are less volatile and the load is higher, leading to fewer opportunities for arbitrage and overall higher costs.

\begin{figure}[bt!]
    \centering
    \includegraphics[width=\linewidth]{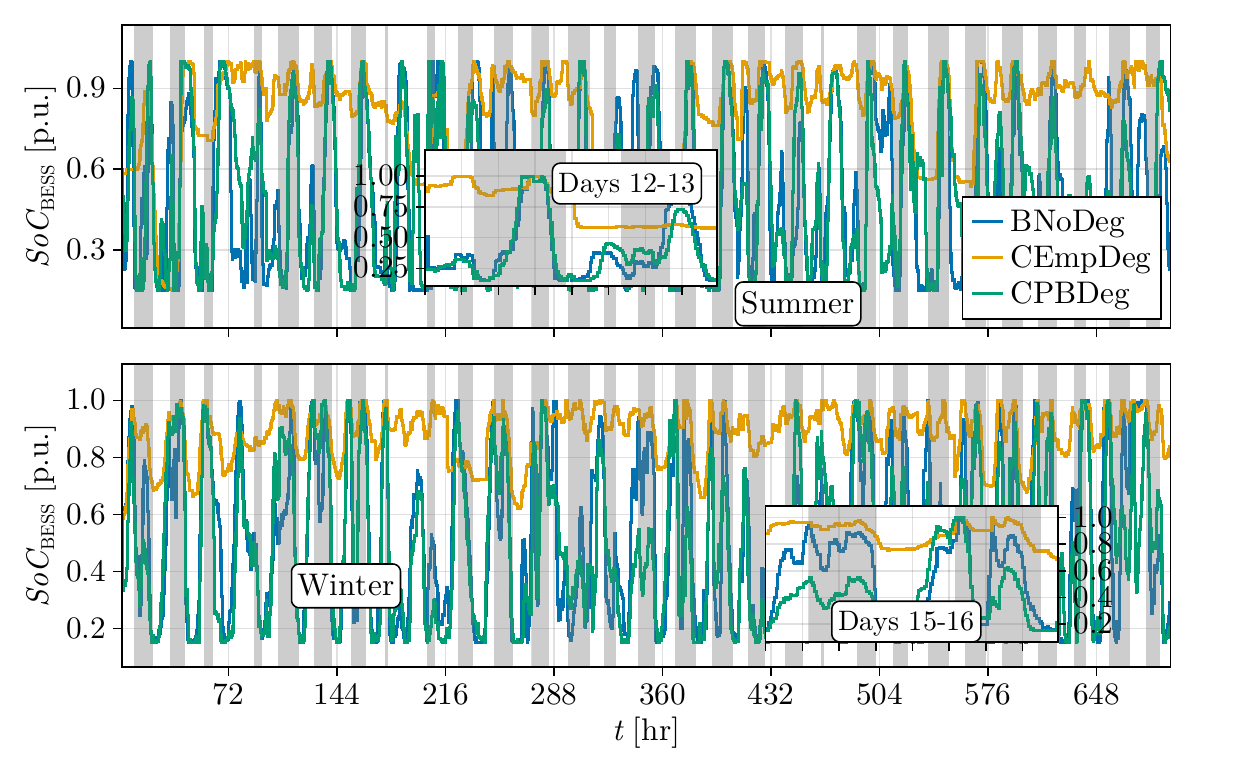}
    \caption{Monthly simulation $SoC_{\textrm{BESS}}$ for summer (top) and winter (bottom).}
    \label{fig:cs1_SoCbess}
\end{figure}

\begin{figure}[bt!]
    \centering
    \includegraphics[width=\linewidth]{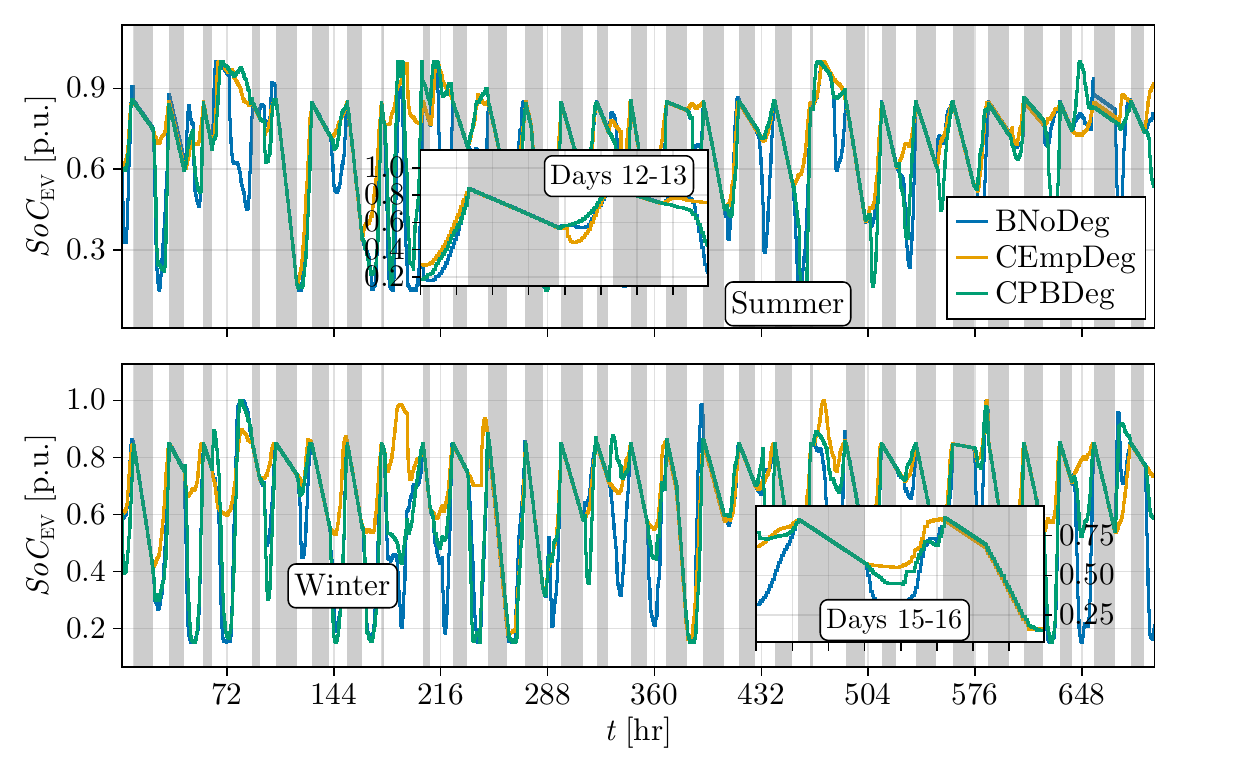}
    \caption{Monthly simulation $SoC_{\textrm{EV}}$ for summer (top) and winter (bottom). Grey bands indicate driving periods.}
    \label{fig:cs1_SoCev}
\end{figure}

\begin{figure}[bt!]
    \centering
    \includegraphics[width=\linewidth]{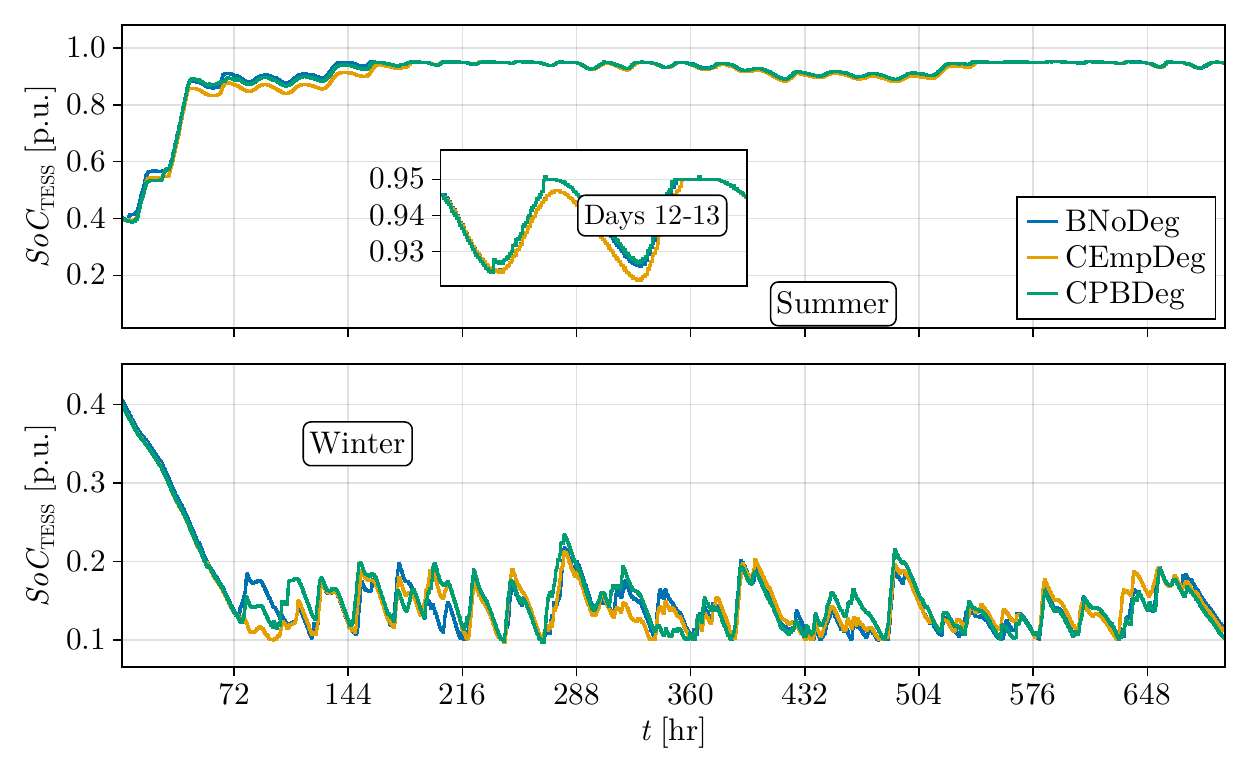}
    \caption{Monthly simulation $SoC_{\textrm{TESS}}$ for summer (top) and winter (bottom).}
    \label{fig:cs1_SoCtess}
\end{figure}

The schedules of the \ac{hess} under the different controllers are summarized in Figs. \ref{fig:cs1_SoCbess}- \ref{fig:cs1_SoCtess}. In the \ac{bess}, Fig. \ref{fig:cs1_SoCbess}, the \textit{BNoDeg} and \textit{CPBDeg} cycle the battery pack more often. This is because \textit{BNoDeg} does not contemplate ageing, and the \textit{CPBDeg} equations relate minimizing $Q_{\text{loss}}$ to $\underline{SoC}$. The \textit{CEmpDeg} controller cycles less frequently, due to its overestimation of $Q_{\text{loss}}$, concentrating the operation around $\overline{SoC}$ to reduce the ageing of the \ac{bess}. This happens both in summer (top) and winter (bottom). Moreover, on many days the price variations are not large enough to afford ageing the \ac{bess}, thus \textit{CPBDeg} chooses to maintain the $SoC$ close to its lower bound. The two highlighted days are days where the \textit{CPBDeg} outperforms the benchmarks and the \ac{bess} prioritizes $C_{\text{grid}}$ over $Q_{\text{loss}}$. 

For the \ac{ev} timeseries comparison, Fig.\ref{fig:cs1_SoCev}, the user's mobility requirement leads to similar timeseries for all the planners. As expected, \textit{BNoDeg} prioritizes short-term revenues and has the most frequent V2G, followed by \textit{CPBDeg} and with \textit{CEmpDeg} usually choosing the highest $SoC_{\text{EV},t}$ of the three. Again, this reflects the dependency of the empirical model, Eq. \ref{eq:empDeg}, with $DoD$. V2G mode suits the \textit{CPBDeg} because it incentivizes discharging the \ac{ev}, minimizing its degradation. This counterintuitive result is because the ageing \ac{pbrom} shows how $Q_{\text{loss}}$ is minimized by $\underline{SoC}$. Since $p_{\text{SoCDep}}$ requires high $SoC^*_{\text{dep}}$, the dynamic combination of them both results in more V2G.

On the thermal side, presented in Fig. \ref{fig:cs1_SoCtess}, the natural periodicity of the carrier is longer due to its thermal efficiencies, size and C-rate. Qualitatively, all three planners take similar decisions independent of the battery model used. This is because the \ac{tess} has a roundtrip efficiency $\eta$ lower than the \ac{ev} and \ac{bess} , and thus a longer optimization window $H$ is needed for the \ac{tess} to show its flexibility value relative to the \ac{bess} or \ac{ev} \cite{Darivianakis2017AManagement, Prat2024HowProblems}.  During the first week the initial $SoC_{\textrm{TESS},0}$ influences costs greatly. After the first week, the \ac{ems} has already steered the buffer to its desired setpoint. This means a high $SoC_{\textrm{TESS},t}$ during summer (high heat generation, low load) and a low setpoint during winter (low heat generation, high load). The high setpoint during summer entails 2 risks: overcharging the \ac{tess} (i.e., activating soft-constraint and curtailing \ac{st}) and not capturing negative prices prices due to past short-sightedness (\ac{tess} starting a day with a high $SoC$). Even though in the summer \textit{CPBDeg} has the smallest \ac{st} curtailment the ageing models have a low overal impact. In other words, integrating ageing models does not change the overall \ac{tess} decisions, since the advanced cell models don't change the fact that $\eta_{b} \geq \eta_{\text{TESS}}$ and a bigger $H$ is needed for the \ac{tess} to present valuable flexibility.

In summary, each storage system follows a distinct use represented in its terminal set. The \ac{ev} follows a soft-tracking penalty $p_{\text{SoCDep}}$, the \ac{bess} follows a flexible daily periodicity, and the \ac{tess} handles medium-term thermal demand. Without the design of the mixed terminal set, the non-convex NLP would not converge to feasible locally optimal points, and thus is a key contribution of this paper.

\begin{table}
\centering
    \begin{tabular}{l l l l l} 
        \hline
        & \multicolumn{2}{c}{$C_{\text{g}}$ [\texteuro]} & \multicolumn{2}{c}{$Q_{\text{loss}}$ [mAh]}\\
         Planner& summer & winter & summer & winter \\
        \hline
         \textbf{BNoDeg} &   55.4&   101.7&  377.3&  400.1\\
         \textbf{CEmpDeg} &   62.3&   115.3&  234.0&  250.8\\
         \textbf{CPBDeg} & 57.6& 109.9& 343.2&351.7\\
    \end{tabular}
\caption{Planner comparison cost summary with weights $w_{\text{grid}}=1,\ w_{\text{loss}}=0.01,\ w_{\text{SoC}}=w_{\text{TESS}}=1000$.}
\label{tab:cs1_costSummary}
\end{table}
The performance of each planner is summarized in Table \ref{tab:cs1_costSummary}. Starting with the grid cost $C_{\textrm{grid}}$, the best performer is the \textit{BNoDeg} in the summer and winter. The worst performer is \textit{CEmpDeg} and the \ac{pb} controller \textit{CPBDeg} stands in the middle. For the total capacity fade $Q_{\textrm{loss}}$, \textit{CEmpDeg} has the lowest degradation while the \textit{BNoDeg} has the highest. The proposed ageing-aware \textit{CPBDeg} has -10\% less degradation than the \textit{BNoDeg}, maintaining a reasonable $C_{\textrm{grid}}$.

\begin{figure}[t!]
    \centering
    \includegraphics[width=\columnwidth]{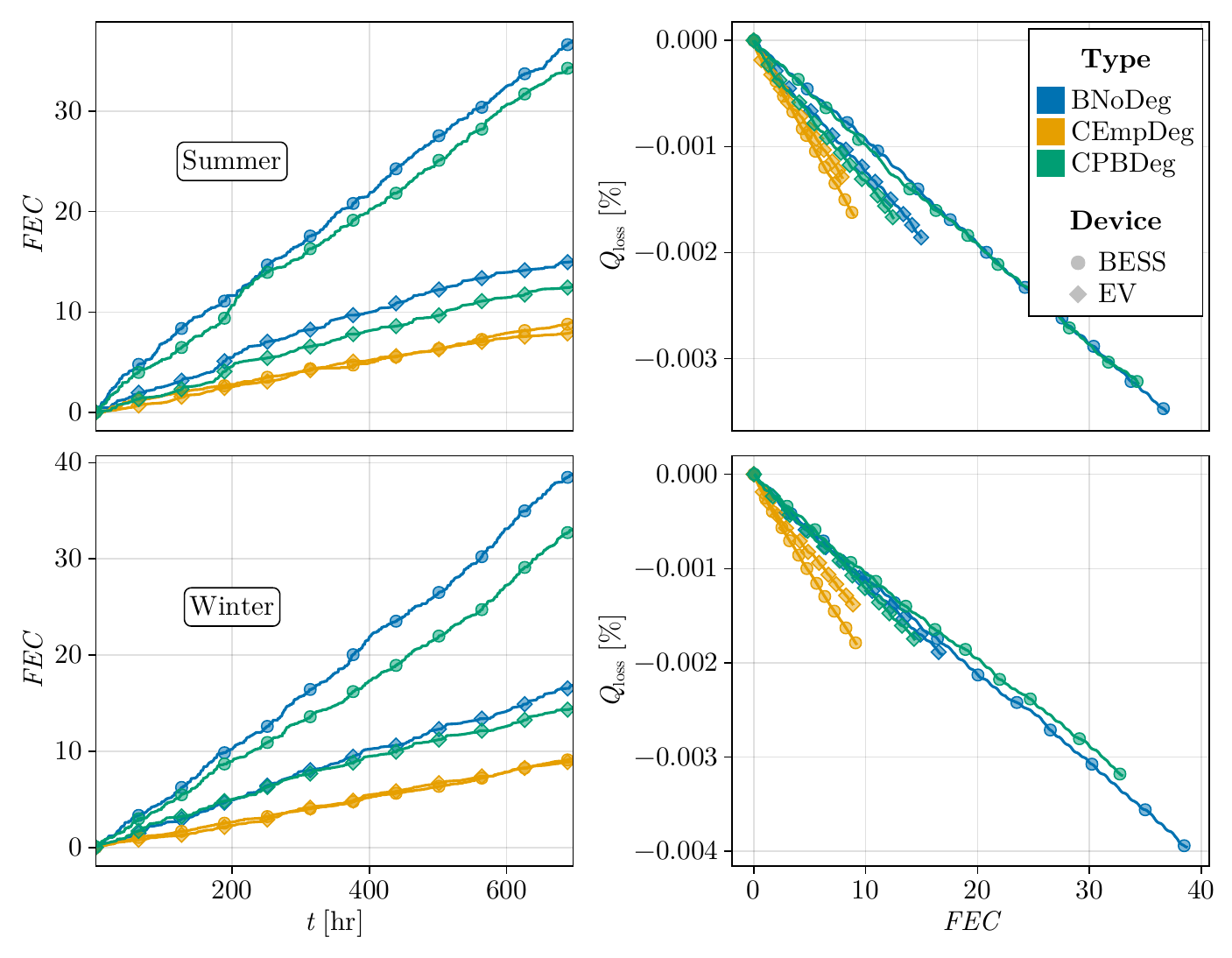}
    \caption{Full equivalent cycles $FEC$ over time (left) and relative capacity fade $Q_{\textrm{loss}}$ over $FEC$ (right).}
    \label{fig:cs1_FECQloss}
\end{figure}

However, just reviewing the objective function $J^{\text{DA/MPC}}$ is not enough. To dive deeper into the total storage usage, a quantitative analysis of the number of cycles done by the $b$ is necessary. Figure \ref{fig:cs1_FECQloss} presents the full equivalent cycles $FEC$ over time $t$, showing that the \textit{BNoDeg} controller has the most $FEC$, followed by \textit{CPBDeg} and with the empirical controller \textit{CEmpDeg} having the least throughput. When analyzing the capacity fade $Q_{\textrm{loss}}$ against the full eq. cycles $FEC$, Fig. \ref{fig:cs1_FECQloss}, it is clear that the relative degradation per cycle ($\frac{\partial Q_{\textrm{loss}}}{\partial FEC}$) of the \textit{CPBDeg} is the smallest of them all. Moreover, its empirical competitor (\textit{CEmpDeg}) has the highest degradation per cycle $\frac{\partial Q_{\textrm{loss}}}{\partial FEC}$, having the most inefficient degradation control. This appears to be a risky strategy due to a lack of consistency across seasons and objectives (minimizing degradation or minimizing grid costs). Lastly, even though the capacity fade is not significant in $T=1$ month, daily optimization can have a significant impact in the long term, as it was shown in \cite{Movahedi2024ExtraMechanism, Cao2020MultiscaleModels, Reniers2021}. As a final note, if the C-rate is increased ($\geq$1) and battery temperature $T_{b,t}$ is not constant, the degradation on a daily basis can be significantly higher.
\begin{figure}[t!]
    \centering
    \includegraphics[width=\columnwidth]{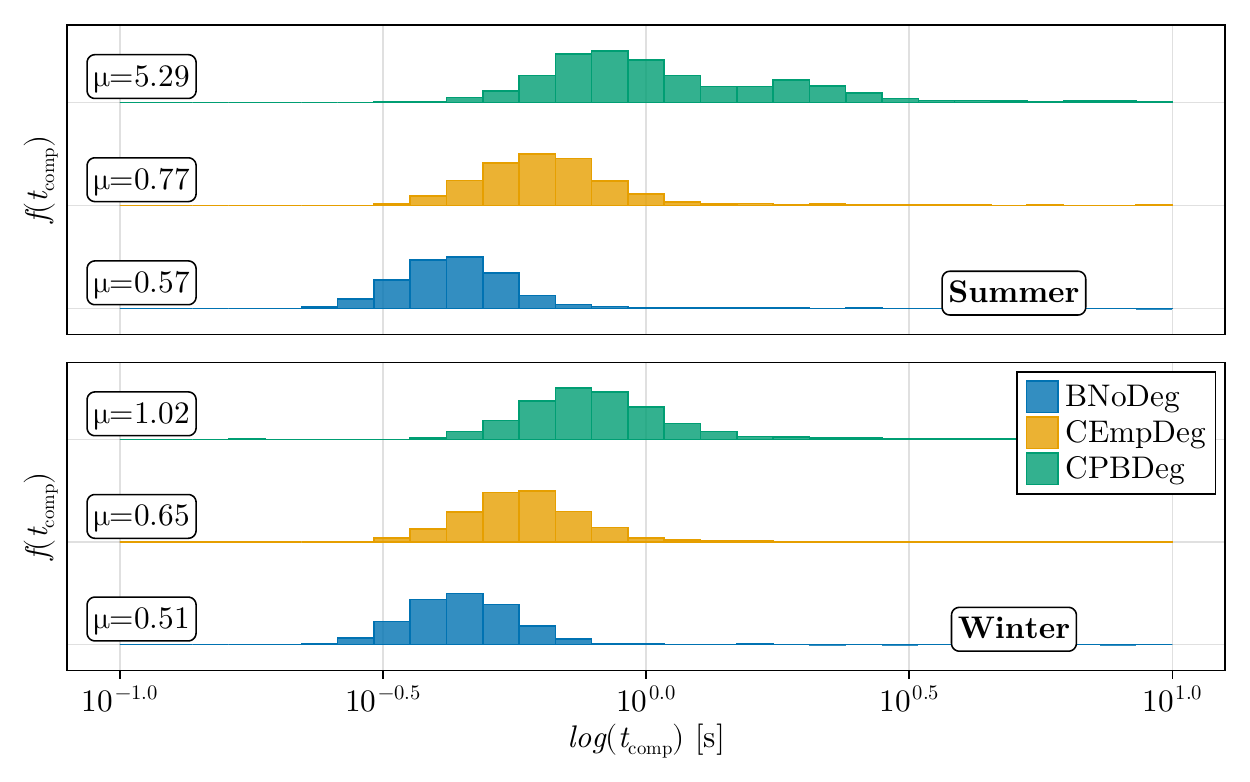}
    \caption{Distributions of computational time $t_{\textrm{comp}}$.}
    \label{fig:cs1_compTime}
\end{figure}
\begin{table}
    \centering
    \begin{tabular}{cccc}\hline
         & \textbf{BNoDeg} & \textbf{CEmpDeg} & \textbf{CPBDeg} \\
         \hline
         \# Variables    & 4416 & 6720 & 6720 \\
         \# Constraints  & 3073 & 5377 & 5377 \\
         \# Linear Eq.   & 2113 & 4033 & 4033 \\
         \# Quad. Eq.    & 768  & 768  & 768  \\
         \# Nonlin. Eq.  & 0    & 384  & 384  \\
         \# Linear Ineq. & 192  & 192  & 192  \\
         \hline
    \end{tabular}
    \caption{Problem size and variables after discretization for each planner.}
    \label{tab:problemSizes}
\end{table}

Finally, the computational time for the different strategies is presented in Fig. \ref{fig:cs1_compTime}.  Each sample is the total computational time it takes to solve Algorithm \ref{alg:CS1-RH}. All \ac{ocp} instances are solved until local optimality, with \textit{BNoDeg} being the closest to its global optima since its a convex QCQP \cite{KNITRO}. Unexpectedly, \textit{BNoDeg} has the lowest and most consistent $t_{\textrm{comp}}$ distribution, i.e., the smallest standard deviation. Both  \textit{CEmpDeg} and \textit{CPBDeg} planners have similar distributions, maintaining overall fast computational time between 1-5s. More importantly, all three distributions overlap with more than 50\% of cumulative probability. The optimization model variables and equations after transcription are presented in terms of their number of variables in Table \ref{tab:problemSizes}. Hence, the increase in modeling accuracy of \ac{pbrom} is not prohibitively expensive when compared to its empirical counterpart. This is to be expected as the empirical ageing model is also non-linear and non-convex. All of these demonstrates the feasibility of implementing the \textit{CPBDeg} in real-time applications.

In summary, this case study shows that:
\begin{itemize}
    \item \textit{CPBDeg} achieves the smallest $\frac{\partial Q_{\textrm{loss}}}{\partial FEC}$ (best ageing control) while maintaining close to best $C_{\text{grid}}$.
    \item The computational cost of \textit{CPBDeg} is not prohibitive when compared to its benchmarks.
    \item \textit{BNoDeg} achieves the smallest $C_{\text{grid}}$ with fast computational times, as expected.
\end{itemize}

\subsection{Case Study II: Managing different cathodes}\label{sse:CS2}
To demonstrate the PB models' flexibility and extended capabilities, the \textit{CPBDeg} controller is tested using two similar battery packs of the same rated capacity $Q$ but using different cells. One is formed with \ac{nmc} cells and the other with \ac{lfp}. Since \ac{lfp} cells have a lower rated capacity of $Q_n=2.3$ Ah and a lower $OCV$, the battery packs have more $N_{s/p}$ to have roughly the same pack-rated capacity as their \ac{nmc} counterparts. The power limits $\overline{P}_{b,t},\ \underline{P}_{b,t}$ are also maintained to make an even comparison. Another weight $w_{\text{loss}}$ is chosen to further highlight differences between cathode chemistries. The driving profile (availability and consumption) are different from the previous Section, thus small deviations in costs are expected.
\begin{figure}[bt]
    \centering
    \includegraphics[width=0.95\linewidth]{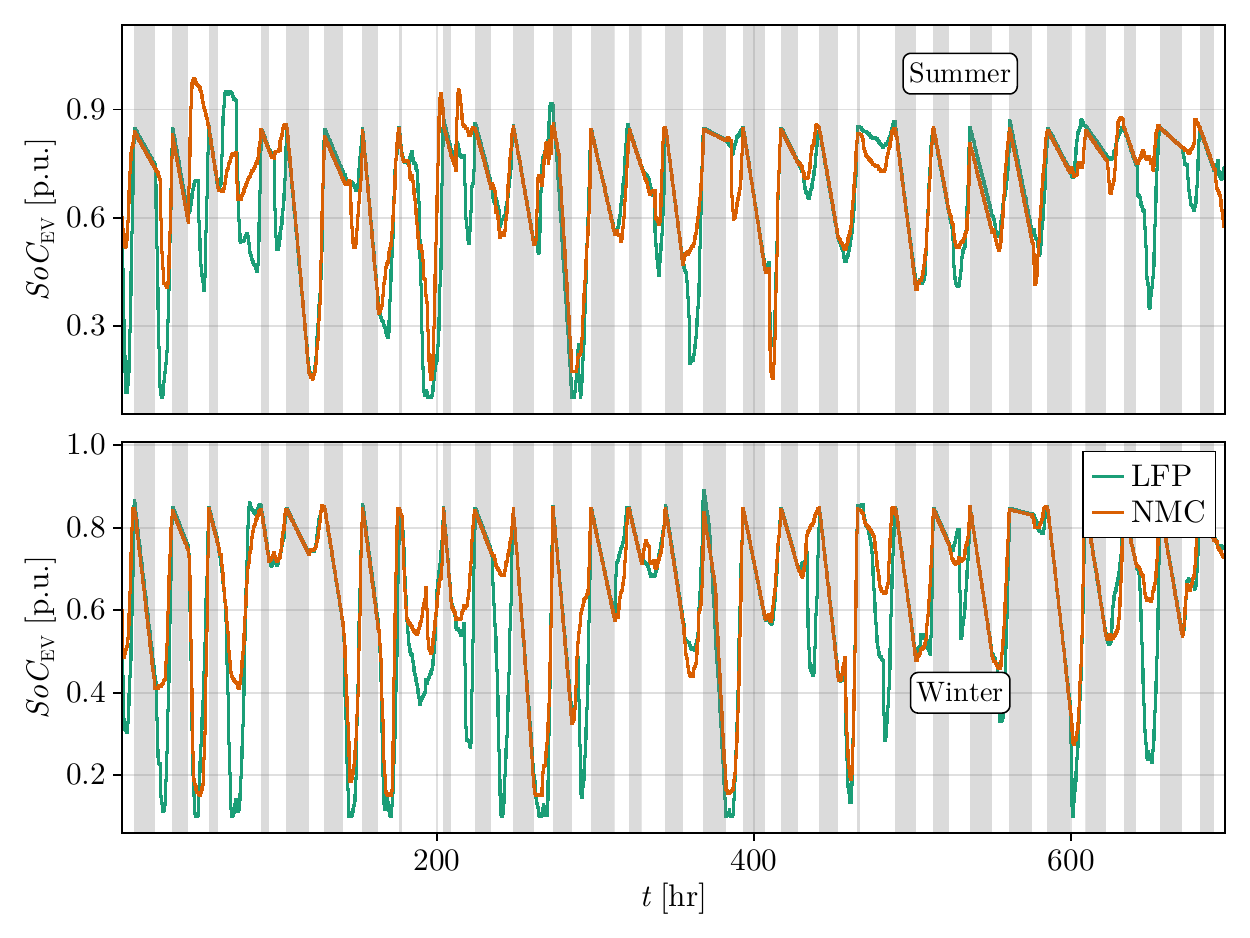}
    \caption{Cathode comparison simulation $SoC_{\textrm{EV}}$. Grey bands indicate driving periods.}
    \label{fig:cs2_SoCev}
\end{figure}

The PB ageing models are suitable for both because they have graphite anodes \cite{Jin2017, Jin2022}. Nevertheless, they have different electrolytes. This is addressed by changing the electrolyte parameters in the model. The same model equations are used, but with different parameter values. This is a great advantage compared to the empirical fits presented in the literature. In the latter, the derived models are prone to overfitting to training conditions, delivering complex non-linear equations that can only be applied to specific chemistries and operating conditions (stationary loads, fixed states).

\begin{figure}[bt]
    \centering
    \includegraphics[width=0.95\linewidth]{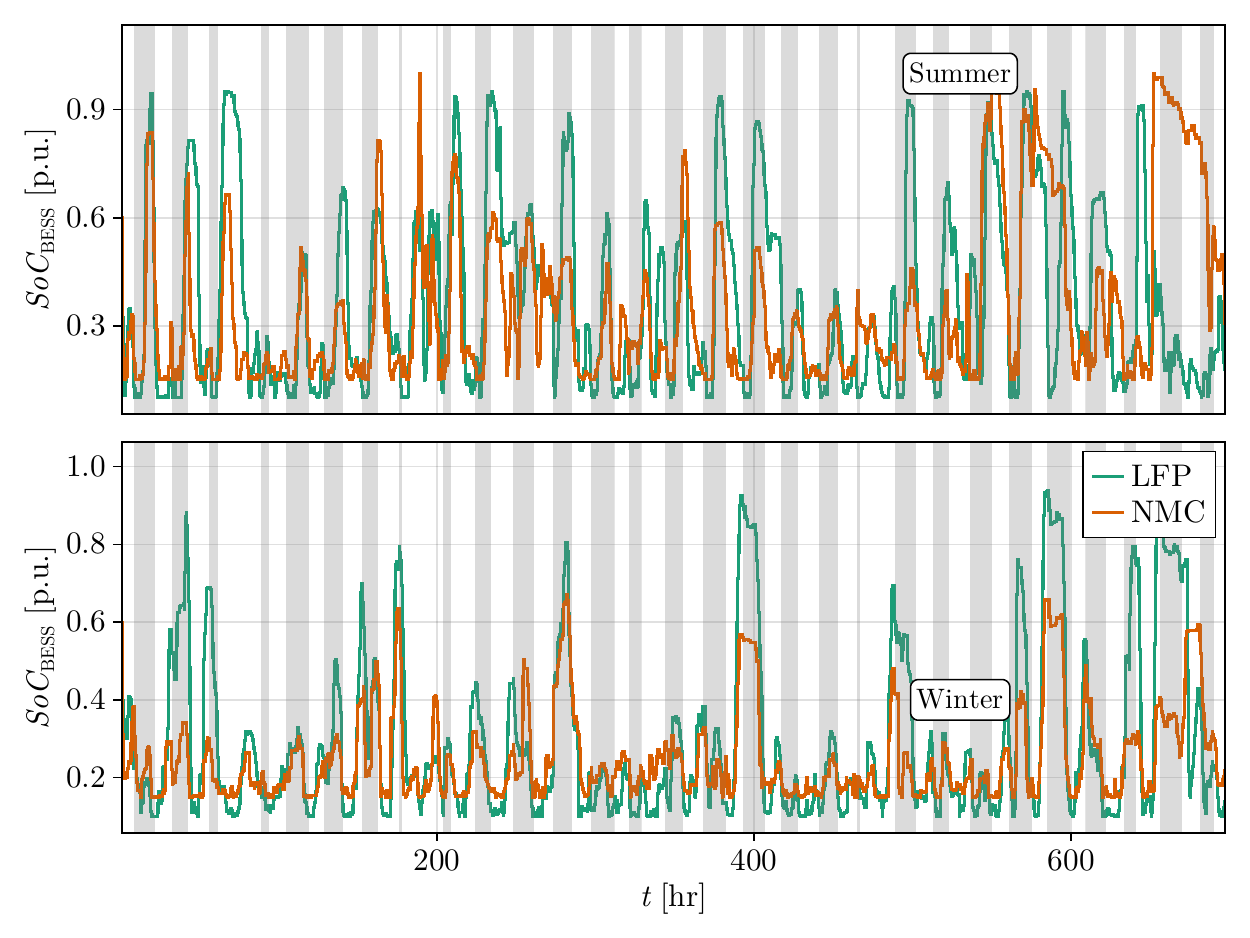}
    \caption{Cathode comparison simulation $SoC_{\textrm{BESS}}$.}
    \label{fig:cs2_SoCbess}
\end{figure}

\begin{figure}[t!]
    \centering
    \makebox[\textwidth][c]{\includegraphics[width=1.\textwidth]{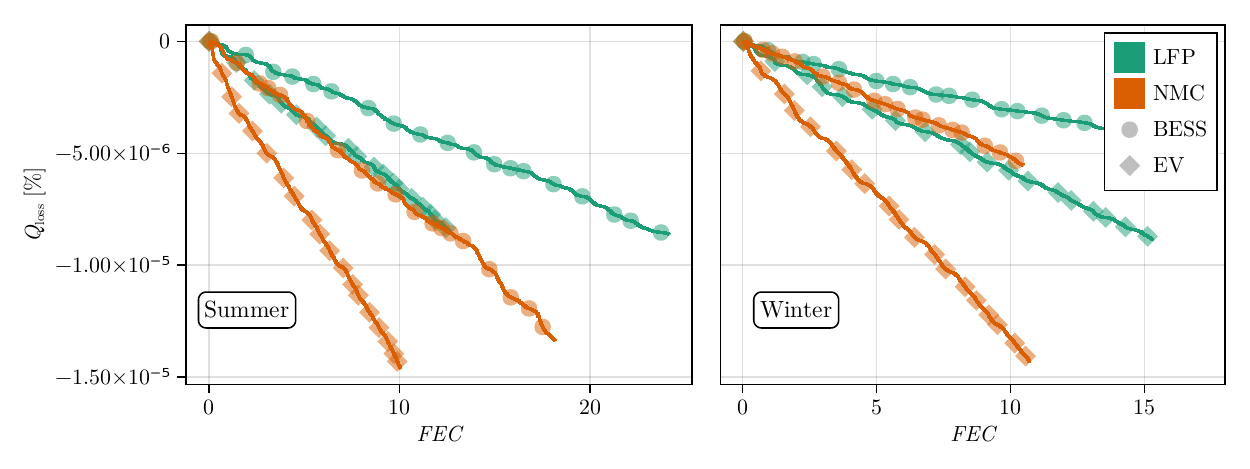}}
    \caption{Case Study 2 - LFP and NMC cells degradation analysis.}
    \label{fig:cs2_QlossFEC}
\end{figure}

The simulation results are presented in Figs. \ref{fig:cs2_SoCev} - \ref{fig:cs2_QlossFEC}.  Starting with the \ac{ev}, Fig. \ref{fig:cs2_SoCev}, the operation is similar except for a few days in summer and winter in which the \ac{lfp} decides to have deeper discharges than its \ac{nmc} counterpart. In other words, the actions in the \ac{nmc} case are more conservative since the \ac{nmc} is more sensitive to the \ac{sei} degradation. Moreover, with the \ac{lfp} cells the \textit{CPBDeg} chooses to do V2G more often than with the \ac{nmc} cell. These leads to the \textit{CPBDeg-LFP} achieving lower grid costs $C_{\text{grid}}$ due to the increased throughput.

Continuing with the ageing analysis Fig. \ref{fig:cs2_QlossFEC} presents the $FEC$ and $Q_{\textrm{loss}}$ results. The overall the $FEC$ are increased with the \ac{lfp} cells. In combination with its lower calendar ageing, represented in the parameter set, the \ac{lfp} packs achieve lower degradation per eq. cycle $\frac{\partial Q_{\textrm{loss}}}{\partial FEC}$ than their \ac{nmc} counterpart. As such, the \textit{CPBDeg-LFP} increases both \ac{bess} and \ac{ev} total throughput ($FEC$) while reducing its degradation  $Q_{\textrm{loss}}$. Overall the change in cathode and consequent electrolyte, reduces the \ac{sei} degradation rate resulting in a more effective degradation control.

Table \ref{tab:cs2_costSummary} presents the summary of performance for both cell types. Overall, the \textit{CPBDeg-LFP} achieves lower total grid costs and ageing. During winter $C_{\textrm{grid}}$ stays the same ($\approx$4\%) while the \ac{lfp} cells has 20\% less degradation. However, in the summer, the \ac{lfp} $C_{\textrm{grid}}$ is 10\% smaller than its \ac{nmc} counterpart, with $C_{\text{grid}}$ similar to \textit{BNoDeg} in Section \ref{sse:CS1} Table \ref{tab:cs1_costSummary}. The \ac{lfp} summer case also has also roughly 20\% less capacity fade than its \ac{nmc} benchmark. This shows that \textit{CPBDeg} rightly exploits its physical information of the system to achieve better performance, both lower grid cost and lower capacity fade.

\begin{table}[bt!]
\centering
    \begin{tabular}{l l l l l} 
        \hline
          $ w_{\text{loss}}=0.1$& \multicolumn{2}{c}{$C_{\text{g}}$ [\texteuro]} & \multicolumn{2}{c}{$Q_{\text{loss}}$ [mAh]}\\
          Cell cathode & summer & winter & summer & winter \\
        \hline
          LFP &   57.9&   112.0&  186.7&  170.5\\
          NMC &   64.7&   116.9&  243.0&  210.2\\
    \end{tabular}
\caption{Cathode comparison cost summary with weights $w_{\text{grid}}=1,\ w_{\text{SoC}}=w_{\text{TESS}}=1000$.}
\label{tab:cs2_costSummary}
\end{table}

\subsection{Case Study III: Managing aged and fresh batteries}\label{sse:CS3}
To demonstrate the flexibility and extended capabilities of the \textit{CPBDeg} planner the scheduler is tested using two battery packs: the first is the fresh battery pack of \ac{nmc} cells of Section \ref{sse:CS1} and the second is the same pack but with cells aged at $SoH=90\%$. Only one benchmark is used: a \textit{BNoDeg} with no $SoH$ update. Thus, the \textit{BNoDeg} \ac{ems} sees a perfectly healthy cell with rated capacity when in reality the battery pack is aged 10\%. There is no empirical benchmark because the only parameter that can be update in it is the initial lifetime $t_{0,b}$, only affecting calendar ageing. The update is based on a 5\%  increase in series resistance $R_{b,0}$ and 10\% decrease of the available Li content $z_{100\%,b}$ and its propagation with the equations of Section \ref{sse:degModel}.

\begin{figure}[t!]
    \centering
    \makebox[\textwidth][c]{\includegraphics[width=1.1\textwidth]{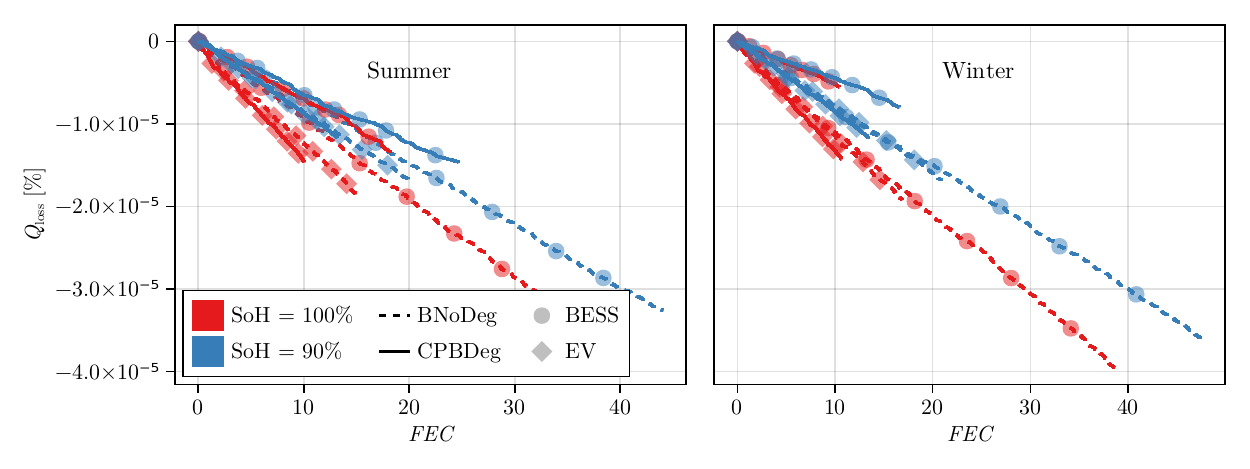}}
    \caption{ Case Study 3 - new and aged cells degradation analysis.}
    \label{fig:cs3_QlossFEC}
\end{figure}

Figure \ref{fig:cs3_QlossFEC} presents the degradation patterns for the 2 planners and different $SoH$. In the aged battery, the share of calendar ageing (against the total) within $i_{\text{SEI},b,t}$ is much smaller and thus the percentual $Q_{\textrm{loss}}$ is almost 25\% smaller than in the new battery packs. In both seasons, the \ac{ev}s patterns are similar for both \textit{BNoDeg} and \textit{CPBDeg}. The change in \ac{ev} trajectories between aged and fresh cells shows a slight increase $FEC$, due to a smaller rated capacity, and a decrease in relative ageing, due to reduced calendar ageing. The impact on the \ac{bess} is more pronounced. In winter, \textit{CPBDeg} has a smaller $\frac{\partial Q_{\textrm{loss}}}{\partial FEC}$ (upper-right hand side of the graph) for both aged and new cells, with \textit{BNoDeg} doing more eq. cycles. In the winter \textit{CPBDeg} still has a smaller $\frac{\partial Q_{\textrm{loss}}}{\partial FEC}$ in the new cell but not in the used cell, where this is achieved by the benchmark controller. In other words, as the battery ages, the predominance of calendar ageing within the SEI layer is diminished. Thus, the link between $i_{b,t}$ and $Q_{\text{loss},b,t}$ becomes stronger, due to the predominance of active material loss and the cyclic component of the SEI layer.

For the current simulation time $T$ of 1 month and average C-rate below 1C, the degradation slope may appear linear to the naked eye. However, in Fig. \ref{fig:cs3_QlossFEC}, the trajectory of the aged battery is also presented, and its average slope is smaller than the new battery degradation slope ($\frac{\Delta Q_{\textrm{loss},b}}{\Delta FEC_b}|_{SoH=90\%}< \frac{\Delta Q_{\textrm{loss},b}}{\Delta FEC_b}|_{SoH=100\%}$). This is due to the nonlinear dependency of $i_{\text{SEI},b,t}$ with $\sqrt t$. Thus, it is clear that even though each trajectory might appear linear, over longer simulation times of several months and years, the dependency is non-linear.

Finally, when adding the costs to the analysis, summarized in Table \ref{tab:cs3_costSummary}, the \textit{CPBDeg} achieves lower capacity fade $Q_{\textrm{loss}}$ than \textit{BNoDeg} across all seasons and $SoH$ (between 30-45\% less). Even improving the total grid cost $C_{\textrm{grid}}$ by 6\% in the summer as the cells degrades. On the other hand, \textit{BNoDeg} worsens its performance as the cells degrade with higher $C_{\textrm{grid}}$ (6-9\%), $Q_{\textrm{loss}}$ and model bias.

In summary, the performance of the proposed \textit{CPBDeg} \ac{empc} controller comparatively improves when using used cells, with respect to its linear \textit{BNoDeg} benchmark. This is because the ageing \ac{pbrom}s encapsulate the fade of calendar ageing ($\sqrt{t}$) and rise of cyclic SEI and AM ($i_{b,t},SoC_{b,t}$) as the battery ages.

\begin{table}
\centering
    \begin{tabular}{l l l l l} 
        \hline
        & \multicolumn{2}{c}{$C_{\text{g}}$ [\texteuro]} & \multicolumn{2}{c}{$Q_{\text{loss}}$ [mAh]}\\
         Planner& summer & winter & summer & winter \\
         \hline
         SoH 100 \%&&&&\\
        \hline
        BNoDeg & 55.39& 101.66  & 377.31  & 400.14  \\
        CPBDeg & 64.68  & 116.86  & 242.92  & 210.18  \\
        \hline
         SoH 90 \%&&&&\\
         \hline
        BNoDeg & 58.66  & 111.31  & 281.11  & 288.31  \\
        CPBDeg & 60.87  & 120.86  & 168.64  & 149.30  \\
        \hline
    \end{tabular}
\caption{Cost and degradation summary with weights $w_{\text{grid}}=1,\ w_{\text{loss}}=0.1,\ w_{\text{SoC}}=w_{\text{TESS}}=1000$.}
\label{tab:cs3_costSummary}
\end{table}

\subsection{Limitations \& Future works}

As it was mentioned in Section \ref{sec:policyDLA} two major assumptions are made regarding the battery packs: (i) moderate C-rates of less than 1C, (ii) constant comfortable battery pack temperature. Both allow us to neglect important factors, such as Li-plating. Temperature control is part of the local primary controls performed by the respective \ac{bms} and no-fast charging outside of the house assumed. Of course, they limit the extension of the presented results to specific days of high heat or low temperatures, since other degradation mechanisms will play a more relevant role. Moreover, as the operation bounds grow more extreme (C-rate and temperatures), the less accurate the models grow even \ac{pb} models. Future research could explore active thermal controls and their impact on the presented degradation control.

The thermal flow-based models used in this paper assume a fixed building temperature decided externally, that produces the exogenous thermal load $P_{\textrm{load}}^{\textrm{th}}$. Another thermal limitation is that efficiencies and conversion factors are assumed linear. These limits the presented work thermal flexibility, since the building's thermal capacity could be used as a passive thermal storage and \ac{hp} consumption prediction could be improved. Both are explored in other works such as \cite{Slaifstein2026SequentialMPC, Alpizar-Castillo2024ModellingHouse}.

A natural limitation of all model-based optimal control methods is that they rely on accurate uncertainty estimation techniques. This means that accurate forecasters are needed to feed belief states $\tilde{B}_t$ of inputs $W_{t+1}$ and state observers/estimators to feedback state measurements $S_{a,t+1}$ and accurately parametrize the transition functions $\tilde{S}^M_t(\cdot)$. The accuracy of both the forecasts $\tilde{B}_{t}$ and device models $\tilde{S}^M_{a,t}(\cdot)$ is a necessary condition for successful real-life policies. The development of battery observers is outside of the scope of this paper but the reader might refer to \cite{Fan2023NondestructivePaths, Li2024NonlinearCharging, Plett2024BatteryMethods, Couto2025DegradationApproach}. The experimental validation of particular models used is outside of the scope of this paper, since we do not want to contribute with a new battery model but with a new dispatch algorithm. The core idea of the present paper is showing that  ageing \ac{pbrom}s, such as the one developed by Jin \cite{Jin2022} or similar, can be integrated to wider EMS/control schemes.

Regarding the periodic constraints of the \ac{bess} and \ac{tess}. Even though the terminal set of the \ac{bess} is more flexible than the standard literature, the design is still arbitrary. It would be interesting to design the terminal conditions as part of a wider control scheme, such as seasonal optimization \cite{Darivianakis2017AManagement}. Not only for the \ac{bess} but for the \ac{tess}, since due to the $H^{\text{DA}}=48$hs, $H^{\text{MPC}}=24$hs its low short-term efficiency leads to discharging and not cycling the \ac{tess}. 

The practical implications for residential consumers are twofold: the short-term grid-cost savings often come at the expense of accelerated battery aging. While the \textit{BNoDeg} and \textit{CEmpDeg} controller can yield lower monthly grid costs ($C_{\text{grid}}$), they do so by aggressively cycling the battery in ways that ignore internal degradation mechanisms. By utilizing the \textit{CPBDeg} control, users can transition from a short-sighted ``bill minimization'' strategy to a ``Total Cost of Ownership'' (TCO) approach. Given the capacity fade cost of approximately 300~\texteuro/kWh, the ability of the \ac{pbrom}-based \ac{ems} to reduce degradation per cycle ($\partial Q_{\text{loss}}/\partial FEC$) suggests that the battery's operational lifespan could be extended by several years. This significantly improves the Net Present Value (NPV) of the residential energy system and reduces the frequency of capital-intensive storage replacements. Naturally, further work is needed to implement similar schemes in real life and until then this work acts as a benchmark to highlight potential savings or practical recommendations. These include, choosing \ac{lfp} battery packs or maintaining batteries close to their $\underline{SoC}_{b,t}$.

\section{Conclusions \& Discussion}
\label{sec:conclusions}

In summary, this paper presents an optimization-based non-linear \ac{empc} for residential multi-carrier energy systems that uses \ac{pbrom} models to integrate battery ageing. Its integration with the thermal and mobility carriers is achieved by designing specific terminal sets for the individual storage systems, depending on their use and technology. The proposed \textit{CPBDeg} control can handle different cathode chemistries as well as batteries in different ageing states. The controller can do this with lower degradation than the benchmarks, by increasing $w_{\text{loss}}$. The \ac{pbrom} integration comes at the expense of slightly higher computational times and grid cost in the case of NMC cells.

In the first case study, it is shown how advanced \ac{pbrom} can be used to reduce battery ageing while maintaining a grid cost comparable to benchmarks, in accordance with the literature for standalone utility-scale applications \cite{Reniers2018, Reniers2019, Reniers2020, Reniers2021, Reniers2023}. Moreover, the \textit{CPBDeg} has the most smallest capacity fade per cycle. This is because the control can effectively relate actions to degradation states. 

In the second case study, the proposed planner is equipped with battery packs of different cathode chemistries, and its performance is compared. Empirical models are cathode-specific and thus can't handle multiple cell chemistries. The \ac{lfp} battery pack has a lower total $Q_{\textrm{loss}}$ as per established knowledge and can achieve lower grid costs $C_{\textrm{grid}}$ than its \ac{nmc} counterpart. Even more so, when considering that the relative cost $c_{\textrm{loss}}$ for both packs was the same, when in reality \ac{lfp} packs have a lower cost than \ac{nmc} packs. Thus, Section \ref{sse:CS2} is a conservative approximation, and \ac{lfp} packs have the potential to enable even lower TCO. The grid savings are achieved just by changing some simple model parameters taken from the literature. This flexibility is an essential feature because it allows the \ac{ems} to exploit any battery pack at hand fully. This directly impacts the choice of the residential users towards \ac{lfp} battery packs, because even without aging-aware controls the overall ageing favors \ac{lfp} packs which is already an industry standard.

The last case study showed how the proposed \ac{ema} handles aged and new batteries seamlessly, even improving performance as the battery ages, improving the summer grid cost by 6\%. The reason is that the \ac{ema} can identify predominant degradation mechanisms from the ageing \ac{pbrom}, exploiting the decrease of SEI importance over time and the rise of active material loss. Its linear counterpart, however, is unaware of the degradation causing the energy costs to increase. Summing up, grid costs can be reduced by using \ac{lfp} battery packs or modifying objective weights, the latter at the expense of higher degradation. Finally, the integration of ageing \ac{pbrom}s has a low impact on the \ac{hp} and \ac{tess}, meaning that the grid cost savings during summer of Section \ref{sse:CS2} are achieved without fundamentally modifying the thermal strategy. In applications where scalability and fast computation are key drivers, decomposition strategies could be explored to exploit this coupling.

On a broader scope, even though relevant challenges are still pending or out of the scope of this paper, this paper aims to show the potential savings enabled by aging-aware control laws in the residential sector. The findings presented here can be easily transferred to other sectors such as commercial or industry, which might have more tools to implement complex \ac{ems}.


For physical setups the proposed physics-based approach requires a non-linear observer to identify internal states and a system identification algorithm to parametrize the models. Adaptive control techniques/ online learning techniques are crucial for scaling implementation. For the case of the \ac{bess} initial tests and model identification can be done offline before start-up, and even offline during operation if historic data is continuously stored.  However, for \ac{ev}s models, parametrization deems a challenge since a previously unknown car may appear or due to unknown driving conditions and profiles. Thus, the development of effective and accurate observers to identify and parametrize \ac{pbrom} online automatically is crucial.


\section{Acknowledgment}

The project was carried out with a Top Sector Energy subsidy from the Ministry of Economic Affairs and Climate, carried out by the Netherlands Enterprise Agency (RVO). The specific subsidy for this project concerns the MOOI subsidy round 2020, FLEXINet project grant number MOOI 32027.

\section{Declaration of generative AI and AI-assisted technologies in the writing process}

During the preparation of this work, the author(s) used ChatGPT and Gemini to help the writing process in favor of clarity and readability. After using this tool/service, the author(s) reviewed and edited the content as needed and take(s) full responsibility for the content of the published article. 

\appendix

\section{Building Battery Models}
\label{sec:appA}

Graphite anode open-circuit voltage \cite{Jin2017, Jin2022}:

\begin{equation}
\begin{aligned}
    OCV_{n,b,t}(z_{b,t}) &= 0.6379 + 0.5416 \cdot e^{-305.5309 \cdot z_{b,t}} \\
    &\quad + 0.044 \cdot \tanh\left(-\frac{z_{b,t} - 0.1958}{0.108}\right) \\
    &\quad - 0.1978 \cdot \tanh\left(\frac{z_{b,t} - 1.0571}{0.0854}\right) \\
    &\quad - 0.6875 \cdot \tanh\left(\frac{z_{b,t} + 0.0117}{0.0529}\right) \\
    &\quad - 0.0175 \cdot \tanh\left(\frac{z_{b,t} - 0.5692}{0.0875}\right)
\end{aligned}
\end{equation}

Table \ref{tab:parametersPBROM} presents the model parameters for the \ac{pbrom}s.

\begin{table}[tbh]
    \centering
        \begin{adjustbox}{width=\textwidth}
            \begin{tabular}{lllll}
                \hline
                \textbf{Parameter} & \textbf{Description} & \textbf{Units}  &\textbf{NMC (SANYO cell)} &\textbf{LFP (A123 cell)}\\ \hline
                $n_{\textrm{SEI}}$ & Number of e$^-$ transferred in SEI side reaction & -  &\multicolumn{2}{c}{2.0}\\ 
                $\lambda$ & Constant $\lambda = \frac{c_s \sqrt{D_s}}{c_p \sqrt{D_p}}$ & -  &\multicolumn{2}{c}{5.51$\times 10^{-5}$}\\ 
                $OCV_s$& OCV of the side reaction & V  &\multicolumn{2}{c}{0.4}\\ 
                $\varepsilon_{\textrm{AM}}$ & Active material volume fraction & -  &\multicolumn{2}{c}{0.552}\\ 
                $R_s$ & Particle radius & m  &7.5$\times 10^{-6}$&5$\times 10^{-6}$\\ 
                $a_s$ & Specific surface area of the anode & m$^{-1}$  &\multicolumn{2}{c}{\( \frac{3 \cdot \varepsilon_{\textrm{AM}}}{R_s} \)}\\ 
                $A_n$ & Active surface area of the anode & m$^2$  &0.105 
                 &0.18\\ 
                $L_n$ & Thickness of anode & m  &50$\times 10^{-6}$&34$\times 10^{-6}$\\ 
                $i_0$ & Exchange current of the intercalation current & A/m$^2$  &\multicolumn{2}{c}{1.5}\\ 
                $k_{\textrm{SEI}}$ & Kinetic rate & 1/$\sqrt{\text{s}}$  &\multicolumn{2}{c}{66.85}\\ 
                $E_{\textrm{SEI}}$ & Activation energy & J/mol  &\multicolumn{2}{c}{39146.0}\\ 
                $\delta_{SEI,0}$& Initial value of the SEI layer thickness & m  &\multicolumn{2}{c}{2.0 $\times 10^{-9}$}\\ 
                $M_{\textrm{SEI}}$ & Molecular weight of the SEI layer & kg/mol  &\multicolumn{2}{c}{0.162}\\ 
                $\rho_{\textrm{SEI}}$ & Density of the SEI layer & kg/m$^3$  &\multicolumn{2}{c}{1690.0}\\ 
                $z_{100\%}$& Full electrode stoichiometry & -  &0.9 
                 &0.81\\ 
                $z_{0\%}$& Empty electrode stoichiometry & -  &0 
                 &0.0176\\ 
                $k_{\textrm{AM}}$ & Kinetic rate & 1/Ah  &\multicolumn{2}{c}{0.0137}\\ 
                $E_{\textrm{AM}}$ & Activation energy & J/mol  &\multicolumn{2}{c}{39500.0}\\ 
                $\beta$ & Tuning parameter & -  &\multicolumn{2}{c}{1.7}\\ 
                $t^{+}_0$ & Transport/transference number & -  &0.363 
                 &0.36\\ 
                $E_{\kappa}$ & Activation energy for $\kappa$ & J/mol  &\multicolumn{2}{c}{34700.0}\\ 
                $E_{D_e}$& Activation energy for $D_e$ & J/mol  &\multicolumn{2}{c}{34700.0}\\ 
                $\kappa_{\textrm{ref}}$& Reference ionic conductivity for $\kappa$ at reference temperature & S/m  &\multicolumn{2}{c}{0.174}\\ 
                $D_{e,\textrm{ref}}$& Reference value for $D_e$ at reference temperature & m$^2$/s  &\multicolumn{2}{c}{7.5$\times 10^{-11}$}\\ 
                $brug$ & Bruggeman coefficient & -  &\multicolumn{2}{c}{3/2}\\ 
                $c_{e,\textrm{avg}}$& Average volume concentration of Li in the electrolyte & mol/m$^3$  &1000&1200\\ 
                $c_{e,\textrm{max}}$& Maximum volume concentration of Li in the electrolyte & mol/m$^3$  &1000&1200\\ 
                $\sigma_n$ & Electronic conductivity & S/m  &100&215\\
                $\varepsilon_{s}$ & Volume fraction of solid electrolyte & -  &0.59  &0.58\\ \hline
            \end{tabular}
        \end{adjustbox}
    
    \caption{Parameter values, descriptions, and units for PBROMs.}
    \label{tab:parametersPBROM}
\end{table}

For the first-order \ac{ecm}s the model parameters are presented in Table \ref{tab:parametersECM}.

\begin{table}[th]
    \centering
    \begin{tabular}{llll}
        \hline
        \textbf{Parameter} & \textbf{Unit} & \begin{tabular}[c]{@{}l@{}}\textbf{NMC} \\ (SANYO cell)\end{tabular} & \begin{tabular}[c]{@{}l@{}}\textbf{LFP} \\ (A123 cell)\end{tabular}\\
        \hline
        $\eta_c$& \% & 99.5&99.9\\
        $Q_{0}$& Ah/cell& 5.29&2.29\\
        $R_0$& $m\Omega$& 28.11&27.01\\
        $\tau_1=R_1 C_1$& s& 2.35&2.13\\
        $R_1$& $m\Omega$& 33.57&26.98\\
        \hline
    \end{tabular}
    \caption{Parameter values, descriptions, and units for ECMs.}
    \label{tab:parametersECM}
\end{table}

\begin{figure}[b!]
    \centering
    \includegraphics[width=0.8\linewidth]{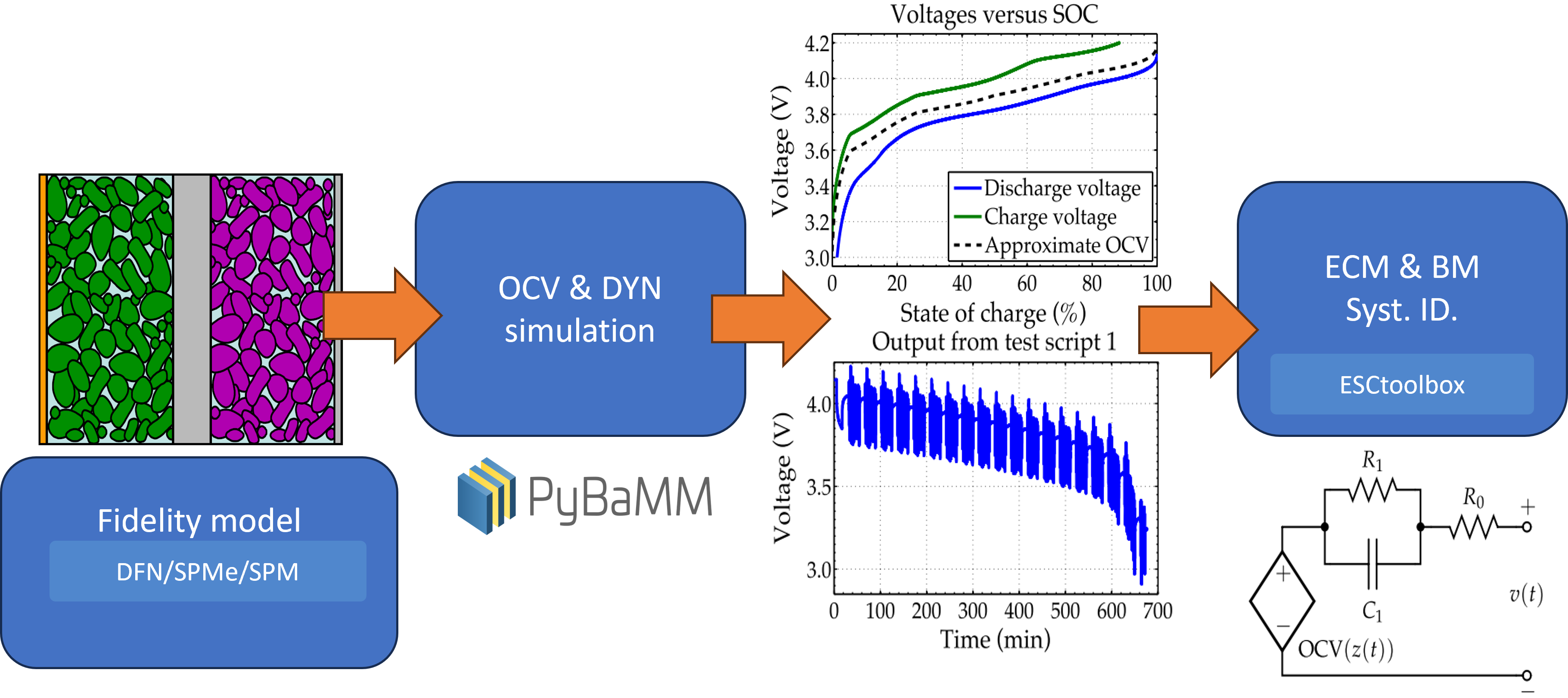}
    \caption{ECM System Identification pipeline used to build planner models.}
    \label{fig:sysIDpipeline}
\end{figure}
In order to test the each of the \ac{ems} decision-making a virtual simulator is necessary to stablish a common ground to compare policies \cite{Powell2022}. A reasonable tool is the \jlinl{PyBaMM}  modelling and simulation library as a digital simulator, which is widely used in the battery community for design, optimization and benchmarking \cite{Sulzer2021PythonPyBaMM}. \jlinl{PyBaMM} has datasets for several real cells, including \ac{nmc} and \ac{lfp} cells \cite{Okane2022, Prada2013Simulations}.

To construct the models used in the optimizer, $\tilde{S}^M_{b,t}(\cdot)$, and avoid using physical cells, one can use a PyBaMM model (compendium of PDEs) and simulate dynamic and open circuit tests as the ones specified in \cite{Plett2016}, and identify an ECM using system identification with the \jlinl{ESCtoolbox}, again following \cite{Plett2016}. The pipeline is presented in Figure \ref{fig:sysIDpipeline}.

To build the simulator, $S_{b,t}^M(\cdot)$, directly using PyBaMM is inconvenient because interfacing \jlinl{Julia} and \jlinl{Python} is possible but hinders execution time. The solution is using the \jlinl{LiiBRA.jl} package \cite{Planden2022}, which builds a state-space PBROM from the SPM or DFN model using Discrete-time Realization Algorithms (DRA). This greatly improves simulation times and allows for detailed simulations with a time resolution of $\Delta t_s=1$s \cite{Planden2022}.

\section{Periodic Condition}
\label{sec:appB}

\begin{figure}[h]
    \centering
    \begin{subfigure}{0.48\textwidth}
        \includegraphics[width=0.9\linewidth]{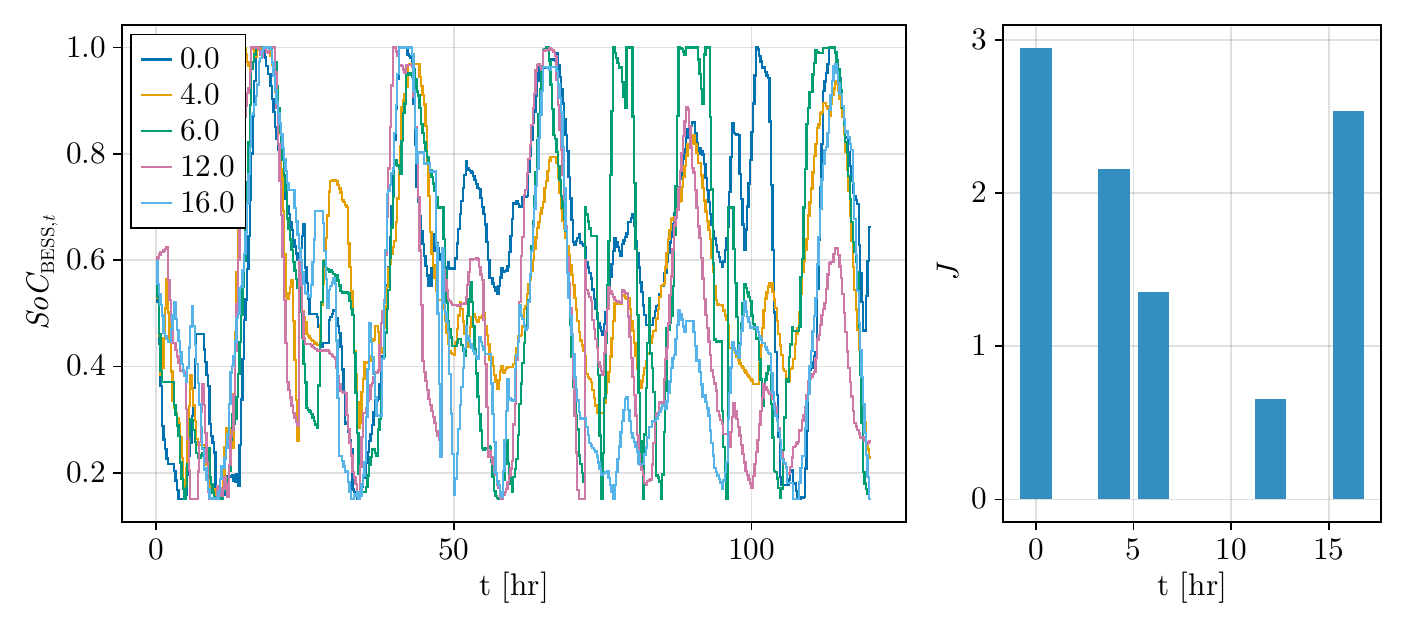} 
        \caption{Summer}
        \label{fig:termCondSummer}
        \end{subfigure}
        \begin{subfigure}{0.48\textwidth}
        \includegraphics[width=0.9\linewidth]{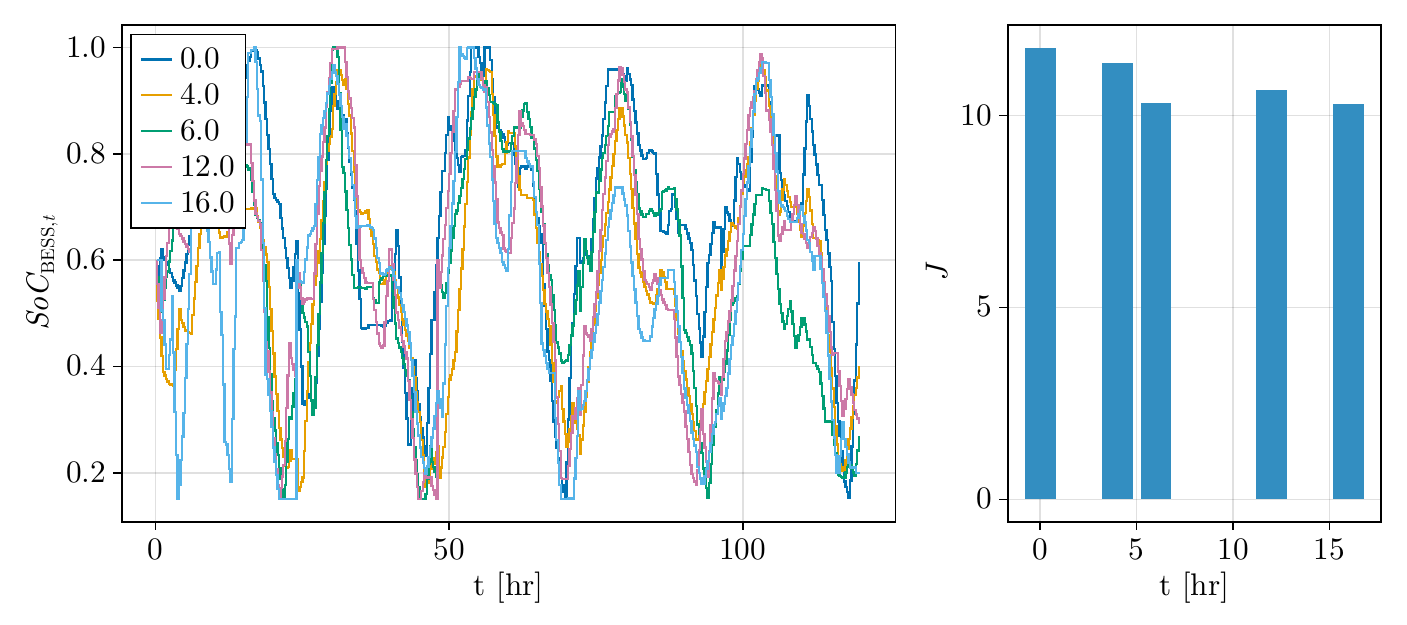}
        \caption{Winter}
        \label{fig:termCondWinter}
    \end{subfigure}
    \caption{BESS State-of-Charge $SoC_{\text{BESS},t}$ and total objective $J$ as a function of terminal condition $t_1$ in (a) summer and (b) winter.}
    \label{fig:termCond2}
\end{figure}

In  Eq. \ref{eq:besspcond} the value of $t_1=6$hs was chosen experimentally through simulations. Figure \ref{fig:termCond2} presents the $SoC_{\text{BESS},t}$ and the total objective $J$ as function of terminal condition $t_1$ for the standard (mean inputs) weeks of summer and winter. In it, the $t_1=6$hs maintains the second place in both weeks, with its closest competitor of $t_1=12$hs is dependent of solar generation which plummets in winter.

\printglossaries
\printnomenclature

\bibliographystyle{elsarticle-num} 
\bibliography{references}
\biboptions{sort&compress}





\end{document}